\documentclass[sigconf]{acmart}

\AtBeginDocument{%
  }

\setcopyright{acmlicensed}
\copyrightyear{2024}
\acmYear{2024}
\acmDOI{XXXXXXX.XXXXXXX}

\acmConference[XX'24']{Make sure to enter the correct 
  conference title from your rights confirmation emai}{XX 25--29,
  2024}{XX, XX}
\acmISBN{978-1-4503-XXXX-X/18/06}

\usepackage{times}
\usepackage{latexsym}

\usepackage[T1]{fontenc}

\usepackage[utf8]{inputenc}

\usepackage{microtype}

\usepackage{hyperref}       
\usepackage{url}            
\usepackage{booktabs}       
\usepackage{amsfonts}       
\usepackage{nicefrac}       
\usepackage{microtype}      
\usepackage{xcolor}         
\usepackage{CJKutf8}
\usepackage{multirow}
\usepackage{setspace}
\usepackage{amsmath}
\usepackage{amsthm}
\usepackage{bbding}
\usepackage{graphicx}
\usepackage{makecell}
\usepackage{tcolorbox}
\usepackage{colortbl}  
\usepackage{array} 
\usepackage{utfsym}

\begin{document}

\title{No Language is an Island: Unifying Chinese and English in Financial Large Language Models, Instruction Data, and Benchmarks}


\author{Gang Hu}
\affiliation{%
  \institution{Yunnan University}
  \city{Kunming}
  \country{China}}
\email{hugang@ynu.edu.cn}

\author{Ke Qin}
\affiliation{%
  \institution{Yunnan University}
  \city{Kunming}
  \country{China}}
\email{kqin@mail.ynu.edu.cn}

\author{Chenhan Yuan}
\affiliation{%
  \institution{The University of Manchester}
  \city{Manchester}
  \country{UK}}
\email{chenhan.yuan@manchester.ac.uk}

\author{Min Peng}
\affiliation{%
  \institution{Wuhan University}
  \city{Wuhan}
  \country{China}}
\email{pengm@whu.edu.cn}

\author{Alejandro Lopez-Lira}
\affiliation{%
 \institution{University of Florida}
 \city{Gainesville}
 \country{America}}
\email{alejandro.lopez-lira@warrington.ufl.edu}

\author{Benyou Wang}
\affiliation{%
  \institution{The Chinese University of Hong Kong}
  \city{Guangdong}
  \country{China}}
\email{wangbenyou@cuhk.edu.cn}

\author{Sophia Ananiadou}
\affiliation{%
	\institution{The University of Manchester}
	\city{Manchester}
	\country{UK}}
\email{sophia.ananiadou@manchester.ac.uk}

\author{Jimin Huang}
\affiliation{%
  \institution{The Fin AI}
  \city{Singapore}
  \country{Singapore}}
\email{jimin.huang@thefin.ai}

\author{Qianqian Xie\textsuperscript{*}}
\affiliation{%
  \institution{The Fin AI}
  \city{Singapore}
  \country{Singapore}}
\email{xqq.sincere@gmail.com}

\renewcommand\footnotemark{}
\authornote{Corresponding author}

\renewcommand{\shortauthors}{Gang Hu et al.}

\renewcommand{\shorttitle}{Unifying Chinese and English in Financial Large Language Models, Instruction Data, and Benchmarks}

\begin{abstract}
    The advancement of Large Language Models (LLMs) has significantly enhanced financial analysis, but their use has been mostly limited to single languages, leaving the potential for bilingual Chinese-English capabilities unexplored. To bridge this gap, we introduce ICE-PIXIU\footnote{\url{https://github.com/YY0649/ICE-PIXIU}}, the first open-source framework that establishes instruction datasets ICE-FIND, fine-tuned LLMs ICE-INTENT, and an evaluation benchmark ICE-FLARE for financial LLMs in both Chinese and English. ICE-FIND includes 18 Chinese-English bilingual specific tasks with 36 datasets and 604k data samples, expanding the breadth and depth of bilingual financial modeling. Harnessing this, we then develop the pioneering Chinese-English bilingual financial LLM ICE-INTENT by fine-tuning the InternLM backbone model. We evaluate our models and existing LLMs using ICE-FLARE, the first comprehensive bilingual evaluation benchmark with 40 datasets covering 10 NLP tasks and 20 bilingual specific tasks. Experimental results demonstrate significant multilingual performance disparities and biases in existing LLMs, with fine-tuning using translation data can enhance LLM's cross-lingual generalization capabilities. ICE-INTENT models outperform SOTA LLMs like GPT-4 in Chinese financial tasks, attributed to diverse linguistic resources, high-quality human-annotated prompts, and strategic instruction fine-tuning.
\end{abstract}





\maketitle

\begin{CJK}{UTF8}{gkai}

\section{Introduction}

As the influence of Chinese investments and consumption grows and English remains the dominant language in global business, effective cross-lingual communication between these languages is increasingly vital~\cite{mohamed2003accounting}. Given the close interaction between the financial markets of China and the United States, bridging linguistic and cultural gaps in the complex financial sector requires specialized tools and models tailored for cross-language business scenarios~\cite{zhang2024d,nie2024survey}.

Inspired by the success of general domain large language models (LLMs), the exploration of financial LLMs as financial cross-language solutions is gradually gaining popularity~\cite{lu2023bbt,xie2023pixiu}. Starting with the broad capabilities of models like OpenAI's ChatGPT~\cite{brown2020language}, the focus has shifted towards more specialized models such as BloombergGPT~\cite{wu2023bloomberggpt}, which tackles the complex nuances of financial language and concepts. 
Subsequently, models like FinGPT~\cite{yang2023fingpt}, InvestLM~\cite{yang2023investlm} and FinTral~\cite{bhatia2024fintral} emerged, concentrating
on English financial tasks. Notably, FinMA~\cite{xie2023pixiu} addresses the challenges of instruction data diversity, marking a significant advancement.

Beyond the efforts in English-centric models, Chinese financial LLMs (FinLLMs) have also made notable advances. Baidu's XuanYuan (轩辕)~\cite{zhang2023xuanyuan} exemplifies this trend, focusing on the intricacies of Chinese financial language. Subsequent models such as DISC-FinLLM~\cite{chen2023disc} and CFGPT~\cite{li2023cfgpt} further enhance capabilities in financial analysis and reasoning. Recently, PanGu-π~\cite{wang2023pangu} enhances a pre-trained model for better performance in financial examinations. 

Although these advancements, there are significant challenges in their application and evaluation in bilingual contexts, especially regarding model development, diversity of instruction datasets, and evaluation methods in English and Chinese. As depicted in Table ~\ref{tab:com}, there remain several issues: (1) All models are tailored exclusively for either English or Chinese, highlighting a significant lack of models that are proficient in both languages; (2) Moreover, most models are not open source, and particularly, nearly all crucial fine-tuning instruction datasets are closed, which will severely impede research in this field. (3) Current evaluation methods often focus on monolingual assessments and fail to comprehensively evaluate bilingual proficiency. Even more concerning is their lack of performance validation with data-out-of-training tasks, leading to potentially inflated evaluation metrics. (4) While existing instruction datasets including financial NLP tasks like classification, extraction, prediction and reasoning, these datasets, particularly for Chinese, show a notable lack of diversity in each task category. (5) Original and corresponding translated datasets are often overlooked in the development of comprehensive bilingual instruction datasets, making it challenging to assess the models' cross-lingual generalisation.

\begin{table*}[htb!]
    \centering
    \footnotesize
    \setlength\tabcolsep{8pt}
    \renewcommand{\arraystretch}{0.8}
    \caption{A comparison of various FinLLMs, including backbone model, parameter size, language proficiency, instruction and evaluation counts on Chinese [zh], English [en], and data-out-of-training [dot] tasks (specific tasks, used only for cross-lingual generalization evaluation in LLMs, not for LLM training), open source (Yes $\checkmark$ and No $\times$) and release date. Task counts for instruction and evaluation across all FinLLMs are based on the specific task types detailed in Table \ref{tab:raw},  with "-" denoting unspecified information in their sources.}
   \begin{tabular}{l|c|c|c|cc|ccc|cc|c}
    \toprule[1pt]
    \multirow{2}{*}{\textbf{FinLLMs}}&\multirow{2}{*}{\textbf{Backbone}}&\multirow{2}{*}{\textbf{Parameter Size}}&\multirow{2}{*}{\textbf{Language}}&\multicolumn{2}{c}{\textbf{Instruction}}&\multicolumn{3}{c}{\textbf{Evaluation}}&\multicolumn{2}{c}{\textbf{Open Source}} & \multirow{2}{*}{\textbf{Release Date}}\\
     &  &  & &  \textbf{zh} &\textbf{en} & \textbf{zh} & \textbf{en} &\textbf{dot} &\textbf{Model}& \textbf{Data}  \\
    \midrule
    BloombergGPT~\cite{wu2023bloomberggpt}&BLOOM~\cite{le2023bloom}&50B&en&0 &0 &0 &0 &5 &$\checkmark$&$\times$&03/30/23\\
    XuanYuan2.0~\cite{zhang2023xuanyuan}&BLOOM~\cite{le2023bloom}&176B&zh&- &0 &6 &0 & 0&$\checkmark$&$\checkmark$&05/19/23\\
    FinMA~\cite{xie2023pixiu}&LLaMA&7/13B&en&0 &5 &0 &5 &0 &$\checkmark$&$\checkmark$&06/08/23\\
    FinGPT~\cite{yang2023fingpt}&ChatGLM~\cite{du2021glm}&7/13B&en&0 &6 &0 &6 &0 &$\checkmark$&$\times$&06/09/23\\
    InvestLM~\cite{yang2023investlm}&LLaMA~\cite{touvron2023llama}&65B&en&0 &4 &0 &4 &0 & $\times$&$\times$&09/15/23\\
    CFGPT~\cite{li2023cfgpt}&InternLM~\cite{team2023internlm}&7B&zh&6 &0 &6 &0 &0 & $\checkmark$&$\times$&09/19/23\\
    DISC-FinLLM~\cite{chen2023disc}&Baichuan~\cite{yang2023baichuan}&13B&zh&8 &0 &8 &0 &0 & $\checkmark$&$\times$&10/23/23\\
    PanGu-π ~\cite{wang2023pangu}&-&1/7B&zh&- &0 &4 &0 &0 & $\times$&$\times$&12/27/23\\
    FinTral~\cite{bhatia2024fintral}&Mistral~\cite{jiang2023mistral} &7B&en& 0 &9 &0 &9 &0 & $\times$&$\times$&02/16/24\\ 
    ICE-INTENT&InternLM~\cite{team2023internlm}&7B&zh, en&12 &6 &12 &6 &3  &$\checkmark$&$\checkmark$&03/10/24\\ 
    \bottomrule[1pt]
    \end{tabular}
    \label{tab:com}
\end{table*}

To tackle this issue, we introduce ICE-PIXIU, the first open-source Chinese-English bilingual framework meticulously designed to bridge the research gap. As detailed in Table~\ref{tab:raw}, ICE-PIXIU has painstakingly curated 40 datasets, incorporating 1,185,076 raw data points, 603,940 fine-tuning instruction data, and 95,091 evaluation data, strategically covering a diverse range of Chinese-English bilingual financial tasks. This diverse array not only underscores our model's commitment to linguistic breadth but also to task-specific depth, ensuring ICE-INTENT's proficiency across various financial scenarios \footnote{Examples of each task are depicted in Appendix Table \ref{tab:exa1} to Table \ref{tab:exa3}}. ICE-PIXIU includes the bilingual open-source multi-task and Chinese-English instructional dataset (ICE-FIND) with 36 datasets spanning 18 specific tasks, the pioneering open-source Chinese-English bilingual evaluation benchmark (ICE-FLARE) with 40 datasets covering 20 specific tasks, and the open-source Chinese-English bilingual financial LLM, ICE-INTENT. 


 To establish a multi-task and bilingual instruction data and evaluation resources, we integrate Chinese and English financial specific tasks from 4 data categories, including (A) Chinese financial datasets, (B) English financial datasets, (C) Chinese translation datasets, and (D) out-of-training financial datasets (as outlined in Section \ref{sec:rawdata}), merging expert-annotated prompts with task-specific data samples. This led to the creation of Chinese and English bilingual financial instruction data called ICE-FIND. To enhance bilingual financial analysis capabilities, we introduce the bilingual financial LLM model, ICE-INTENT, achieved by fine-tuning the InternLM-7B backbone model with ICE-FIND. For evaluation, we further construct the bilingual ICE-FLARE benchmark by including test set from ICE-FIND, and introducing the data-out-of-training (DOT) task along with 3 unseen specific tasks and 4 unseen datasets. The inclusion of DOT task in the benchmark aims to assess the model's generalization capabilities across various financial contexts and tasks.


We assess ICE-INTENT and baseline LLMs with the ICE-FLARE bilingual benchmark. Our analysis revealed three key findings: 1) Current LLMs, such as GPT-4, show a notable performance gap in Chinese financial tasks, highlighting disparities across languages. 2) The ICE-INTENT models excel in bilingual financial analysis, outperforming established models like GPT-4 in Chinese financial tasks. This success emphasizes the pivotal role of fine-tuning instructions and utilizing target language and high-resource datasets to enhance model capabilities. 3) Of particular note, fine-tuning LLMs with translated data not only bridges cross-lingual financial tasks but also significantly boosts the model's performance in English tasks, demonstrating the beneficial effects of cross-lingual data transfer. 

Our contributions can be summarized as follows: \textbf{1) Bilingual Capability Proficiency}: ICE-INTENT, a bilingual financial LLM, showcases exceptional bilingual proficiency in English and Chinese, essential for navigating cross-lingual financial Q\&A scenarios. \textbf{2) Diverse Chinese Financial Datasets}: ICE-PIXIU enhances training and performance by integrating the Chinese classification, extraction, reasoning, and prediction NLP tasks, surpassing existing FinLLMs in task diversity and data scale. \textbf{3) Professionally Annotated Prompts}: ICE-PIXIU provides a diverse, high-quality, expert-annotated set of prompts applied with similar fine-tuning instructions to enhance understanding of financial tasks. \textbf{4) Integration of English-to-Chinese Translation Data}: The framework extends its capabilities by incorporating the English-to-Chinese translation datasets, bolstering bilingual training and cross-lingual generalization. \textbf{5) Cross-Lingual Evaluation with ICE-FLARE}: ICE-PIXIU introduces ICE-FLARE, a rigorous cross-lingual evaluation benchmark that ensures consistent model performance in diverse linguistic contexts. \textbf{6) Open-Source Contribution}: With its open-access approach, ICE-PIXIU offers its models, data, and evaluation resources to the research community, fostering collaborative advancement in financial NLP.

\section{Related Work}
\label{sec:related}

\noindent\textbf{Financial Dataset Resource} 
Financial pre-trained language models (FinPLMs) previously focus on single-task datasets but now most FinLLMs prioritize financial specific-task data collection. For instance, BloombergGPT~\cite{wu2023bloomberggpt} integrates general and financial text to train large 50B LLMs. FinLLMs such as FinGPT~\cite{yang2023fingpt,zhang2023instruct} curate English instruction datasets for classification, extraction, and generation NLP tasks. PIXIU ~\cite{xie2023pixiu}  provide open-source, multi-task and multi-modal instruction data covering various financial scenarios. BBT-Fin~\cite{lu2023bbt} Corpus aggregates Chinese financial data including corporate announcements, research reports, social media, and more, while XuanYuan 2.0~\cite{zhang2023xuanyuan} provides a hybrid tuning dataset. Cornucopia's~\cite{Cornucopia2023} 26M instruction dataset covers application areas such as insurance and stocks. Most fine-tuning datasets are tailored for monolingual tasks, which limits their coverage.

\noindent\textbf{Financial Language Models} 
FinPLMs and FinLLMs build on backbone models like BERT~\cite{devlin2018bert} and LLaMa~\cite{touvron2023llama} with large-scale financial texts. FinBERT~\cite{araci2019finbert} was the first FinPLM with an openly released corpus, later improved FinBERT~\cite{yang2020finbert} incorporating larger communication datasets. Models like FLANG~\cite{shah2022flue} combine BERT and ELECTRA~\cite{clark2020electra}. Chinese-specific PLMs like Mengzi~\cite{zhang2021mengzi} and FinT5~\cite{lu2023bbt} address unique financial tasks. BloombergGPT introduces the outstanding FinLLM, along with successors like FinGPT~\cite{yang2023fingpt} and FinMA~\cite{xie2023pixiu} excelling in specific data collection and supervised fine-tuning using backbone models Baichuan~\cite{yang2023baichuan} and InternLM~\cite{team2023internlm}. However, monolingual FinLLMs struggle to offer consistent cross-lingual finance services.

\noindent\textbf{Financial Evaluation Benchmarks}
The first benchmark assesses classification, boundary detection, and question-answering in English, while FLARE~\cite{xie2023pixiu} extends this to include time-series reasoning. BBT-CFLEB~\cite{lu2023bbt} is the first Chinese benchmark with 5 financial NLP tasks, while FinCUGE ~\cite{chen2023disc} focuses on 9 financial NLP tasks on social media and short news. FinEval~\cite{zhang2023fineval} assesses LLMs through multiple-choice questions on finance, economics, and accounting. CFBenchmark ~\cite{lei2023cfbenchmark} evaluates text processing abilities in recognition, classification, and generation, encompassing totaling 8 tasks, and CGCE~\cite{zhang2023cgce} includes 200 general and 150 specific professionally diverse QA tasks. FinBen~\cite{xie2024finben} serves as an extensive benchmark with 36 datasets over 24 financial tasks. However, these benchmarks lack support for cross-lingual and out-of-domain tasks, posing a challenge for evaluating the generalization ability of FinLLMs.

\section{Method}

\subsection{ICE-FIND: Bilingual Financial Instruction Dataset}
We present ICE-FIND, a Chinese-English bilingual financial instruction dataset, detailing the raw data sources and the instruction construction process using high-quality prompts by human evaluation. Unlike current resources, ICE-FIND emerges as the first instruction data tailored for bilingual FinLLMs, showcasing originality.

\subsubsection{Raw Data Integration}
\label{sec:rawdata}

ICE-FIND is built from open-source data, covering a range of specific tasks in real-world financial scenarios. It stands out by utilizing expert-curated, high-quality sources, offering cost-efficiency over self-instruct data generation methods~\cite{wang2022self} without usage restrictions, and providing varied data formats and multilingual support for flexible task adaptation. As shown in Table \ref{tab:raw}, it curated 5 data types-Data-Label-from-Classification (DLC), Data Labels from-Extraction (DLE), Data-Text-by-Translation (DTT), Data-Text-with-English (DTE), and Data-Out-of-Training (DOT). These categories support 10 financial NLP tasks such as Classification (CLS), Extraction (EXT), Reasoning (REA), Prediction (PRE), Generation (GEN), and Translation (TRA) across 12 Chinese and 6 English specific tasks. Compared with similar studies~\cite{wang2023finvis,chen2023disc} respectively focus on PRE, as well as CLS and REA NLP task, ~\cite{lei2023cfbenchmark,li2023cfgpt} emphasizes CLS and EXT NLP task. Our raw data comprises 40 different datasets covering almost all financial NLP tasks, including CLS, EXT, REA, PRE, GEN, and TRA. 

Our proposed ICE-FIND consists of four datasets as follows: (A) Chinese financial datasets, (B) English financial datasets, (C) Chinese translation datasets, and (D) out-of-training financial datasets. 

\begin{table*}[htb!]
 \centering
    \footnotesize
     \setlength\tabcolsep{6pt}
         \renewcommand{\arraystretch}{0.8}
        \caption{Detailed information on the raw data used for constructing instruction and evaluation dataset of the ICE-PIXIU framework.}
        \begin{tabular}{c|c|c|c|c|c|c|c|c|c}
        \toprule[1pt]
        \textbf{Language} & \textbf{Data Type} & \textbf{NLP Task} & \textbf{Specific Task} & \textbf{Dataset} & \textbf{Raw} & \textbf{Instruction} & \textbf{Evaluation} & \textbf{Data Source}  & \textbf{License} \\ 
        \midrule
        \multirow{25}{*}{ZH} & \multirow{10}{*}{DLC} &\multirow{9}{*}{ZH-CLS} & \multirow{2}{*}{FinSA} & FE & 18,177 & 18,177 & 2,020 & social texts & Public \\ 
        ~ & ~ & ~ & ~ & StockB & 9,812 & 9,812 & 1,962 & social texts & Apache-2.0 \\ 
        \cmidrule(lr){4-10}
        ~ & ~ & ~ & \multirow{2}{*}{FinSM} & BQC & 120,000 & 110,000 & 10,000 & bank service logs & Public \\ 
        ~ & ~ & ~ & ~ & AFQMC & 38,650 & 38,650 & 4,316 & online chat service & Apache-2.0 \\
           \cmidrule(lr){4-10}
        ~ & ~ & ~ & \multirow{2}{*}{FinNC} & NL & 7,955 & 7,955 & 884 & news articles & Public \\ 
        ~ & ~ & ~ & ~ & NL2 & 7,955 & 7,955 & 884 & news articles & Public \\ 
        \cmidrule(lr){4-10}
        ~ & ~ & ~ & FinNJ & NSP & 4,499 & 4,499 & 500 & social texts & Public \\ 
        ~ & ~ & ~ & FinAS & FinevalF & 1,115 & 1,115 & 222 & financial exam & Apache-2.0 \\ 
        ~ & ~ & ~ & FinRE & RE & 14,973 & 14,973 & 1,489 & news, entity pairs & Public \\
          \cmidrule(lr){3-10}
        ~ & ~ & ZH-PRE & FinSP & StockA & 14,769 & 14,769 & 1,477 & news,historical prices & Public \\ 
        \cmidrule(lr){2-10}   
        ~ & \multirow{7}{*}{DLE} & \multirow{6}{*}{ZH-EXT} & FinQA & QA & 22,375 & 22,375 & 2,469 & QA pairs of news & Public \\ 
        \cmidrule(lr){4-10}
        ~ &~ & ~ & FinER & CNER & 1,685 & 1,685 & 337 & financial reports & Public \\ 
         \cmidrule(lr){4-10}
        ~ &~ & ~ & \multirow{4}{*}{FinED} & 19CCKS & 156,834 & 14,674 & 2,936 & social texts & CC BY-SA 4.0 \\ 
        ~ &~ & ~ & ~ & 20CCKS & 372,810 & 45,796 & 9,159 & news, reports & CC BY-SA 4.0 \\ 
        ~ &~ & ~ & ~ & 21CCKS & 8,000 & 7,000 & 1,400 & news, reports & CC BY-SA 4.0 \\ 
        ~ &~ & ~ & ~ & 22CCKS & 109,555 & 59,143 & 11,829 & news, reports & CC BY-SA 4.0 \\ 
        \cmidrule(lr){3-10}
        ~ & ~ & ZH-GEN & FinTS & NA & 32,400 & 32,400 & 3,600 & news, announcements & Public \\
      \cmidrule(lr){2-10}
        ~ & \multirow{8}{*}{DTT} & \multirow{8}{*}{ZH-TRA} & \multirow{2}{*}{FinSA} & CFPB & 4,845 & 4,838 & 970 & economic news & MIT license \\ 
        ~ &~ & ~ & ~ & CFiQA-SA & 1,173 & 1,143 & 233 & ews headlines,tweets & MIT license \\ 
          \cmidrule(lr){4-10}
        ~ &~ & ~ & \multirow{3}{*}{FinSP} & CACL18 & 27,056 & 2,555 & 511 & tweets, historical prices & MIT license \\ 
        ~ &~ & ~ & ~ & CBigData22 & 7,167 & 798 & 159 & tweets, historical prices & MIT license \\ 
        ~ & ~ & ~ & ~ & CCIKM18 & 4,970 & 431 & 86 & tweets, historical prices & MIT license \\ 
          \cmidrule(lr){4-10}
        ~ &~ & ~ & FinHC & CHeadlines & 102,708 & 10,256 & 2,051 & news headlines & MIT license \\ 
          \cmidrule(lr){4-10}
        ~ &~ & ~ & \multirow{2}{*}{FinQA} & CEnQA & 8,281 & 668 & 133 & earnings reports & MIT license \\ 
        ~ &~ & ~ & ~ & CConFinQA & 12,594 & 1,189 & 237 & earnings reports & MIT license \\ 
         \midrule
        \midrule
        \multirow{15}{*}{EN} & \multirow{11}{*}{DTE} &\multirow{5}{*}{EN-CLS} & \multirow{2}{*}{FinSA} & FPB & 4,845 & 4,845 & 970 & economic news & CC BY-SA 3.0 \\ 
        ~ &~ & ~ & ~ & FiQA-SA & 1,173 & 1,173 & 235 & news headlines,tweets & Public \\ 
        \cmidrule(lr){4-10}
        ~ &~ & ~ & FinHC & Headlines & 11,412 & 102,708 & 20,547 & news headlines & CC BY-SA 3.0 \\ 
        \cmidrule(lr){4-10}
        ~ & ~ & ~ & \multirow{2}{*}{FinCC} & German & 1,000 & 1,000 & 200 & credit records & CC BY-SA 4.0 \\ 
        ~ & ~ & ~ & ~ & Australian & 690 & 690 & 139 & credit records & CC BY-SA 4.0 \\
         \cmidrule(lr){3-10}
        ~ &~ & \multirow{3}{*}{EN-PRE} & \multirow{3}{*}{FinSP} & ACL18 & 27,053 & 27,053 & 3,720 & tweets, historical prices & MIT License \\ 
        ~ &~ & ~ & ~ & BigData22 & 7,164 & 7,164 & 1,472 & tweets, historical prices & Public \\ 
        ~ &~ & ~ & ~ & CIKM18 & 4,967 & 4,967 & 1,143 & tweets, historical prices & Public \\ 
         \cmidrule(lr){3-10}
        ~ &~ & EN-EXT & FinER & NER & 609 & 609 & 98 & financial agreements & CC BY-SA 3.0 \\ 
       \cmidrule(lr){3-9}
        ~ &~ & \multirow{2}{*}{EN-REA} & \multirow{2}{*}{FinQA} & EnQA & 8,281 & 8,281 & 1,147 & earnings reports & MIT License \\ 
        ~ & ~ & ~ &  & ConFinQA & 3,458 & 12,594 & 1,490 & earnings reports & MIT License \\ 
         \cmidrule(lr){2-10}
        ~ & \multirow{4}{*}{DOT} & \multirow{4}{*}{EN-DOT} & FinER & FINER-ORD & 1,075 & - & 1,075 & news articles & CC BY-SA 4.0 \\ 
        \cmidrule(lr){4-10}
        ~ & ~ & ~ & \multirow{2}{*}{FinTS} & ECTSUM & 495 & - & 495 & earning call transcipts & Public \\ 
        ~ & ~ & ~ & ~ & EDTSUM & 2,000 & - & 2,000 & news articles & Public \\ 
        \cmidrule(lr){4-10}
        ~ & ~ & ~ & FinDC & FOMC & 496 & - & 496 & FOMC transcripts & CC BY-SA 4.0 \\ 
    \bottomrule[1pt]
    \end{tabular}
    \label{tab:raw}
\end{table*}

\textbf{(A) Chinese Financial Datasets.}
    Our primary focus is on Chinese language proficiency, with the cleaned and processed datasets can be categorized into two data types: DLC and DLE, which encompass classification (ZH-CLS), extraction (ZH-EXT), prediction (ZH-PRE) and generation (ZH-GEN) financial NLP tasks.  
    
    For \textbf{ZH-CLS}, it covers essential tasks like sentiment analysis (FinSA), semantic matching (FinSM), news classification (FinNC), negative judgment (FinNJ), answer selection (FinAS)  and relationship extraction (FinRE).  1) \textbf{\textit{Sentiment Analysis}}. It is the process of analyzing and labeling the sentiment expressed in financial texts as "Negative(消极)","Neutral(中性)" and "Negative(消极)"~\cite{sohangir2018big}. This task integrates the FE~\cite{lu2023bbt} dataset from Stockbar and Xueqiu websites, along with the StockB~\cite{kuroneko5943} dataset that covers finance across ten domains.  2) \textbf{\textit{Semantic Matching}}. It determines whether financial texts or sentences are semantically similar ("是(YES)" or "否(NO)"), using two datasets: the Bank Question Corpus (BQC) ~\cite{chen2018bq} from Chinese bank customer logs and the Ant Financial Question Matching Corpus (AFQMC) ~\cite{xu2020clue} from the Alipay competition. 3) \textbf{\textit{Negative Judgment}}. It focuses on identifying finance and economics-related negativity, as opposed to general financial sentiment analysis. Leveraging the FinNSP dataset~\cite{lu2023bbt} sourced from social media, it uses "是(YES)" to label data with negative information and "否(NO)" for those without. 4) \textbf{\textit{News Classification}}. It involves classifying news into two or more class label, such as "股票市场(stock market)", "外汇(foreign exchange)", "宏观经济(macroeconomics)", and more. This task employs the FinNL~\cite{lu2023bbt}  dataset, sourced from financial news websites, containing both binary (NL) and multi-class (NL2) labeled data.  5) \textbf{\textit{Answer Selection}}. It entails selecting the correct option ("A", "B", "C", "D") within the financial context to evaluate financial reasoning and decision-making skills. AntGroup's FinEval ~\cite{zhang2023fineval,fineva2023} includes 4,661 questions covering finance, economy, accounting, and certification. We compile the available financial subject data into FinEvalF.  6) \textbf{\textit{Relationship Extraction}}. It aims to provide the types of connections, dependencies, and associations between entities. We process the FinRE dataset~\cite{lu2023bbt} into RE, comprising financial news articles with head-tail entity pairs, spanning 44 relationship categories (e.g., "合作(cooperation)", "持股(shareholding)", etc.).
    
    For \textbf{ZH-EXT}, it includes the question answering (FinQA), named entity recognition (FinER), and event detection (FinED) tasks.  1) \textbf{\textit{Question Answering}}. It aims to extract precise answers from financial text for questions about pertinent events. We gather a dataset named QA~\cite{lu2023bbt} to differentiate its name from the English FinQA dataset. 2) \textbf{\textit{Named Entity Recognition(NER)}}. It aims to identify entity names and types labeled as individuals ("个人(PER)"), organizations ("组织(ORG)"), or locations ("地点(LOC)") in financial text. We utilize the the comprehensive NER dataset~\cite{jia2020entity} called CNER.  3) \textbf{\textit{Event Detection}}. It involves recognizing and comprehending specific incidents beyond industry categorization. This task integrates the 19CCKS, 20CCKS, 21CCKS, and 22CCKS datasets ~\cite{ccks1922} from the China Conference on Knowledge Graph and Semantic Computing (CCKS).  Unlike CFGPT~\cite{li2023cfgpt} that focuses on detecting 254 different event types across four categories, our task additionally involves identifying causal relationships between events ("原因事件类型(Cause event type)"→"结果事件类型(Result event type)").
    
    For \textbf{ZH-PRE}, it centers on the \textbf{\textit{Stock Prediction}} (FinSP) task, which entails categorizing stocks into three categories ("表现不佳(under-performing)", "跑赢大盘(outperforming the market)", "中性(neutral)") by combining stock data and financial news. This task utilizes the StockA dataset~\cite{zou2022astock}, showcasing stock-specific news and factors within China's A-shares market. It stands out from similar datasets~\cite{xu2018stock, zhou2021trade} by delivering thorough data on individual stocks.
    
    For \textbf{ZH-GEN}, it encompasses the \textbf{\textit{Text Summarization}} (FinTS) task, aiming to condense complex financial information into concise summaries. The FinNA dataset~\cite{lu2023bbt}, referred to as NA for this task, consists of financial news articles along with their abstracts.  

\textbf{(B) English Financial Datasets.}
    For English, we integrate datasets from PIXIU~\cite{xie2023pixiu} tailored for English-centric FinLLMs. It includes five tasks in financial NLP: classification (EN-CLS), prediction (EN-PRE), extraction (EN-EXT), reasoning (EN-REA). 

    For \textbf{EN-CLS}, it consists of three tasks: sentiment analysis (FinSA), headline classification (FinHC) and credit classification (FinCC). 1) \textbf{\textit{Sentiment Analysis}}. It uses the Financial Phrase Bank (FPB)~\cite{malo2014good} and FiQA-SA~\cite{maia2018} datasets. FPB comprises a human-annotated phrase bank for financial and economic texts, while FiQA-SA mainly comes from social data. 2) \textbf{\textit{Headline Classification}}. It utilizes the Headlines~\cite{sinha2021impact} dataset, comprising gold-related news headlines from 2000 to 2019 annotated with nine categories like "price up”, "price down", "price stable”, "future price", and "asset comparisons". 3) \textbf{\textit{Credit Classification}}. It aims to assess a customer’s credit status using a set of attributes and assigning a credit label of "good" or "bad". This task employs the German~\cite{german144} and Australian~\cite{australian143} datasets.  
    
    For \textbf{EN-PRE}, it treats \textbf{\textit{Stock Prediction}} (FinSP) as a ternary classification task ~\cite{soun2022accurate}, utilizing social media, financial news, and historical stock price data to forecast the close price change rate. The classification criteria are as follows: 1) Close price change rate > 0.50\% is labeled "Rise"; 2) Close price change rate < -0.50\% is labeled "Fall"; 3) Other cases are labeled as "Hold". Three common datasets used are ACL18~\cite{xu2018stock}, CIKM18~\cite{wu2018hybrid}, and BigData22~\cite{soun2022accurate}. 
    
    For \textbf{EN-EXT}, it centers on the \textbf{\textit{NER}} (FinER) task. We use the CoNLL-2003~\cite{alvarado2015domain} dataset, named NER, for annotating financial agreements from U.S. SEC filings into location ("LOC"), organization ("ORG"), person ("PER"), and miscellaneous ("MISC") entities.
    
    For \textbf{EN-REA}, it differs from the Chinese \textbf{\textit{Question Answering}} (FinQA) task in that it integrates stock time series and tables to infer answers. Two datasets used are FinQA~\cite{chen2021finqa} and ConvFinQA~\cite{chen2022convfinqa}. FinQA consists of expert-annotated Q\&A pairs derived from financial reports of S\&P 500 companies within the FinTabNet dataset~\cite{zheng2021global}. ConvFinQA extends FinQA to allow for multi-turn conversations.

\textbf{(C) Chinese Translation Datasets.}
    To enhance the model's cross-lingual capabilities and explore the performance influence of incorporating translated data, we create new English-to-Chinese Translation (ZH-TRA) Financial NLP datasets, covering 4 specific financial tasks: sentiment analysis (FinSA), headlines classification (FinHC), question answering (FinQA) and Stock Prediction (FinSP). 
    
    Specifically, we utilize ChatGPT 3.5-Turbo-1106 to translate the 8 English datasets involved in these specific tasks into Chinese. Notably, we perform complete translations (including training, validation, and testing sets) for the only-text modality datasets within the FinSA and FinHC tasks. Conversely, for FinSP and FinQA, which incorporate additional sequence and tabular modalities, we focus on translating the only testing sets. In FinSA, the English datasets FPB and FiQA-SA are translated into Chinese as CFPB and CFiQA-SA, while in FinHC, the English Headlines dataset is translated into Chinese as CHeadlines. In FinQA, we specifically tailor translation prompts for the EnQA and ConvFinQA English datasets to preserve key information like tables and stock names, creating the Chinese CEnQA and CConvFinQA datasets. In FinSP, we utilize regular expressions to extract text containing task prompts and stock descriptions. With manually crafted translation prompts, we translate the text descriptions while retaining stock details like stock tickers and historical stock prices, resulting in the Chinese datasets named CACL18, CCIKM18, and CBigData22, respectively.  
    
    To balance data integrity and machine translation quality, we eliminate the flaw data as follows: 1) Expert review of translation discrepancies in language structures and specialized terms (Participant details as outlined in Section \ref{sec:instruction}) . 2) ChatGPT-turbo's constraint of 4096 tokens, unable to process excessively long texts with time-series data. An example of the translation method used for the Bigdata22 dataset in FinSP is depicted in Appendix Figure \ref{fig:tra}.

\textbf{(D) Out-of-Training Financial Datasets.}
    A factor overlooked by nearly all existing methods is the introduction of out-of-training datasets (not used for model fine-tuning) to explore their robustness and generalization ability. Thus, we introduce the \textbf{EN-OFT} financial NLP task including NER (FinER), text summarization (FinTS) and hawkish-dovish classification (FinDC), which are not used during the training process. 1) \textbf{\textit{Named Entity Recognition}}. It utilizes the FINER-ORD ~\cite{shah2023finer} dataset, which provides three labels for entities such as "PER", "LOC", and "ORG". 2) \textbf{\textit{Text Summarization}}. Two datasets ECTSUM~\cite{mukherjee2022ectsum} and EDTSUM~\cite{zhou2021trade}, are utilized for this task. ECTSUM involves extracting key sentences from texts, while EDTSUM focuses on generating fitting news headlines through abstractive summarization. 3) \textbf{\textit{Hawkish-dovish Classification}}. Unlike FinSA, it involves classifying sentences from monetary policy texts into a "hawkish" or "dovish" stance using the FOMC ~\cite{shah2023trillion} dataset.

\subsubsection{Instruction Construction}
\label{sec:instruction}

Based on raw data, we further build b\textbf{I}lingual \textbf{C}hinese-\textbf{E}nglish \textbf{F}inancial \textbf{I}nstructio\textbf{N} \textbf{D}ataset (ICE-FIND), covering 20 specific tasks and 40 datasets with a total of 604k data samples, as detailed in Table \ref{tab:raw}. Building ICE-FIND requires integrating diverse prompts into raw data. We convene domain experts to write 30-40 unique prompts for each dataset. Examples of prompts and their translations are shown in Appendix Figure \ref{tab:tra}. However, it is crucial to acknowledge that expert-authored prompts may not consistently yield high-quality results. Thus, we undertake a manual evaluation of the prompts to ensure their effectiveness. 

Specifically, the evaluation panel consists of 8 graduate students and 2 undergraduate students, all proficient in both Chinese and English, with backgrounds in finance domain. Each participant is provided with a questionnaire comprising 3-4 prompts (across 10 rounds) for quality assessment, covering three metrics: accuracy, naturalness, and informativeness. The questionnaire options include: "Strongly Disagree(0)", "Disagree(1)", "Agree(2)", "Strongly Agree(3)". Participants test the prompts with a randomly selected LLM on the ChatALL~\cite{ChatALL2024} platform to rate each prompt from 0 to 3 across all metrics. ChatALL contains various LLMs, such as Bing Chat~\cite{bingchat2024} and ChatGLM~\cite{du2021glm}.  We retain the prompts with an average score above 2, ensuring adaptability to most cases. Examples of human evaluation are presented in Appendix Table \ref{tab:eval1}-\ref{tab:eval2}. 

To address prompt variation in LLMs, we adopt two strategies. For English tasks, we pair each data with all prompts except for FinQA and ConvFinQA. For Chinese tasks, given the larger raw data, one prompt is randomly selected for matching each data. Finally, we convert the raw datasets into instruction-tuning samples by restructuring expert prompts, input texts, and responses. Instruction tuning samples adheres to the structured template as follows.

\begin{table}[htbp]
\centering
    \small
    \begin{tabular}{|p{8cm}|}
    \hline 
    \rowcolor{blue!2}   \{\textbf{Id}:\textcolor{blue}{[data\_id]},\\
    \rowcolor{blue!2}   \makebox[0.5cm]{}\{\textbf{Instruction}:\textcolor{blue}{[prompt]}\makebox[0.5cm]{}\textbf{Input}:\textcolor{blue}{[text]}\makebox[0.5cm]{}\{\textbf{Output}:\textcolor{blue}{[response]}\}\\
    \rowcolor{blue!2}   \}\\
    \hline 
    \end{tabular} 
\end{table}

which integrates human-designed instructions with input texts and their corresponding responses. \textcolor{blue}{[data\_id]} represents the ID of each data, \textcolor{blue}{[prompt]} is the task-specific prompt, \textcolor{blue}{[text]} refers to the input text, \textcolor{blue}{[response]} denotes the expected answer.

\subsection{ICE-INTERN: Bilingual Financial Large Language Model}

Upon evaluating bilingual performance of various backbone LLMs on ICE-FLARE, we explore using InternLM-7B as backbone model with ICE-FIND fine-tuning to develop a bilingual FinLLM named ICE-INTERN-7B. To assess the impact of various data types on fine-tuning ICE-INTERN, we gradually incorporate the DLC, DLE, DTT and DTE datasets during the fine-tuning process, creating ICE-INTERN variants named ICEdlc-7B (ZH), ICEdle-7B (ZH), ICEdtt-7B (ZH), and ICEfull-7B (ZH \& EN), respectively. To optimize memory usage and reduce training times, we introduce an advanced fine-tuning technique called QLoRA ~\cite{hu2021lora}. The same fine-tuning hyper-parameters set for all models are described follows. Specifically, the training sequences are uniformly cut to a length of 2048 tokens. Then, we use the AdamW~\cite{kingma2014adam} optimizer with an initial learning rate of 5e-5 on a weight decay 1e-5 for all parameters. Additionally, the warm-up steps are set to 1\% of the total training steps. Finally, we fine-tune the model with 1 epoch in batch size of 24 on 8 NVIDIA HGX A100 SXM4 GPUs. It is worth noting that our ICEdtt-7B and ICEfull-7B models can be used for language-specific ablation studies, while combining the ICEdle-7B and ICEdtt-7B models enables a deeper exploration of how task-specific Chinese translated datasets can enhance the model's bilingual capabilities.

\subsection{ICE-FLARE: Bilingual Financial Evaluation Benchmark}

\begin{figure}[htb!]
    \centering
    \includegraphics[width=2.5in]{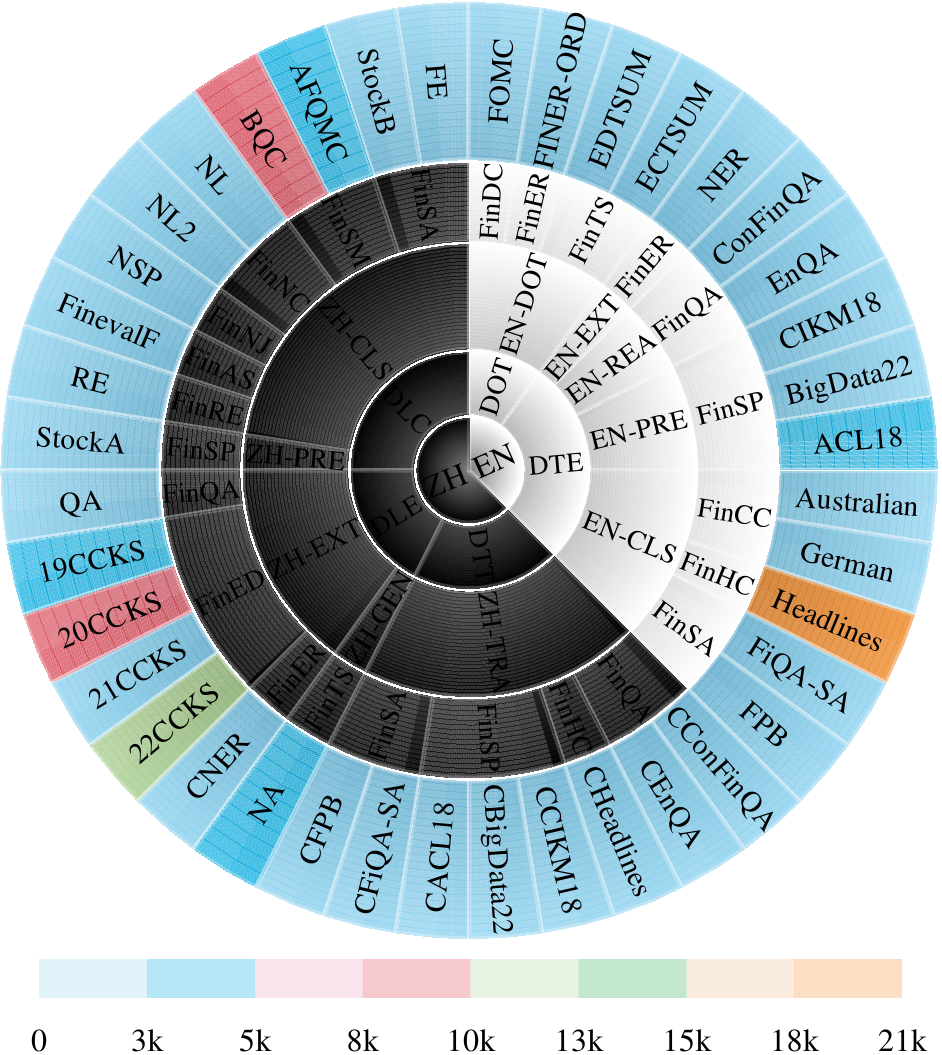}
    \caption{A sunburst chart showing ICE-FLARE distribution across language, data types, NLP and specific tasks, and datasets.}
    \label{fig:map}
\end{figure}

Based on ICE-FIND, we propose the ICE-FLARE bilingual evaluation benchmark. ICE-FLARE can evaluate the  bilingual capabilities of LLMs in the financial domain. The comparison of existing benchmarks ( [a] CFBenchmark~\cite{lei2023cfbenchmark}, [b] FinCUGE~\cite{chen2023disc}, [c] Fineval~\cite{zhang2023fineval}, [d] CGCE~\cite{zhang2023cgce}, [e] CFinBench~\cite{nie2024cfinbench}, [f] FLUE~\cite{shah2022flue},[g] FLARE~\cite{xie2023pixiu}, [h]FinanceBench~\cite{islam2023financebench}, [i]BizBench~\cite{koncel2023bizbench}, [j]FinBen~\cite{xie2024finben}) with ICE-FLARE is detailed in Table \ref{tab:eva}. Unfortunately, they are all designed exclusively for either Chinese or English. From Table \ref{tab:eva}, it is evident that ICE-FLARE, which introduces bilingual capabilities, significantly surpasses these benchmarks on task diversity and data scalability. Despite FinCUGE having a higher number of financial evaluation tasks, the difference in dataset size compared to ICE-FLARE is nearly 8 times. These public benchmarks have provided valuable references and inspiration for constructing the test set, making ICE-FLARE more comprehensive in terms of task coverage, metric diversity, and sample quantity. However, they still face challenges in evaluating the bilingual capabilities in Chinese-English and lack DOT tasks (e.g., FinDC, FinST, FinER) to test the generalization ability of the FinLLMs. In conclusion, ICE-FLARE provides a bilingual evaluation benchmark that covers a sufficient range of diversity tasks and more comprehensive measurement metrics, which is crucial for improving the performance and linguistic capabilities of FinLLMs. An overview of ICE-FLARE, as shown in Figure \ref{fig:map}.

\begin{table}[htb!]
    \centering
    \scriptsize
    \setlength\tabcolsep{0.5pt}
    \renewcommand{\arraystretch}{0.8}
    \caption{The detail information about bilingual benchmark ICE-FLARE. "Cover" indicates whether (Yes $\checkmark$ and No $\times$) these public monolingual financial evaluation benchmarks ([a] CFBenchmark, [b] FinCUGE, [c] Fineval, [d] CGCE, [e] CFinBench, [f] FLUE, [g] FLARE, [h] FinanceBench, [i] BizBench, [j] FinBen) can cover the specific task within ICE-FLARE.}
    \begin{tabular}{cc|cccc|cccccccccc}
    \toprule[1pt]
\multicolumn{2}{c|}{\multirow{2}{*}{\textbf{Specific Task}}} &\multicolumn{2}{c}{\multirow{2}{*}{\textbf{Metric\quad Language}}} &   \multirow{2}{*}{\textbf{Data}}  & \multirow{2}{*}{\textbf{Test}} &  \multicolumn{5}{c}{\textbf{Chinese}} & \multicolumn{5}{c}{\textbf{English}}   \\                                                    
     &  &   &                          &                            &             &     [a]&[b]&[c]&[d] &[e] &[f] &[g]&[h] &[i] &[j]\\
\midrule
\multirow{6}{*}{\makecell{Sentiment \\Analysis}} &  \multirow{6}{*}{\rotatebox[origin=c]{90}{FinSA}}    & \multirow{8}{*}{\makecell{F1\\Accuracy}}           & \multirow{4}{*}{zh} & FE         & 2,020  &\multirow{6}{*}{$\checkmark$}&\multirow{6}{*}{$\checkmark$}&\multirow{6}{*}{$\times$}&\multirow{6}{*}{$\times$} &\multirow{6}{*}{$\times$}&\multirow{6}{*}{$\checkmark$}&\multirow{6}{*}{$\checkmark$}&\multirow{6}{*}{$\times$}&\multirow{6}{*}{$\times$}&\multirow{6}{*}{$\checkmark$}  \\
                                         &      &                              &                     & StockB     & 1,962	                 \\
                                         &      &                             &                     & CFPB       & 970                 \\
                                         &      &                              &                     & CFiQA-SA   & 233                 \\
\cmidrule(lr){4-6}
                                         &      &                              &    \multirow{2}{*}{en} & FPB        &  970                 \\
                                        &      &                              &                     & FiQA-SA    &  235                \\
\midrule
\makecell{Semantic\\Matching}& \rotatebox[origin=c]{90}{FinSM}    &\makecell{F1\\Accuracy} & zh & \makecell{Corpus\\AFQMC} &  \makecell{10,000\\4,316}  & $\times$&$\times$&$\times$&$\times$&$\times$ & $\times$& $\times$&$\times$& $\times$& $\times$ \\                   
\midrule
\makecell{News\\Classification} & \rotatebox[origin=c]{90}{FinNC}   &\makecell{F1\\Accuracy} & zh & \makecell{NL\\NL2} & \makecell{884\\884}  &$\checkmark$ &$\checkmark$&$\times$&$\times$ &$\times$ &$\times$&$\times$&$\times$& $\times$& $\times$ \\
\midrule
\makecell{Negative\\Judgment} &   \rotatebox[origin=c]{90}{FinNJ}   & \makecell{F1\\Accuracy}      & zh        & NSP        & 500    &$\times$ &$\checkmark$ &$\times$ &$\times$  &$\times$ &$\times$ & $\times$ &$\times$ & $\times$& $\times$               \\
\midrule
\makecell{Answer\\Selection} &\rotatebox[origin=c]{90}{FinAS}      & \makecell{F1\\Accuracy}          & zh      & FinevalF  &  222   &$\times$ &$\checkmark$  &$\checkmark$   &$\times$ &$\checkmark$ &$\times$  &  $\times$ &$\times$  &$\checkmark$& $\times$             \\
\midrule
\multirow{7}{*}{\makecell{Stock\\ Prediction}}  &\multirow{7}{*}{\rotatebox[origin=c]{90}{FinSP}}      & \multirow{7}{*}{\makecell{F1\\Accuracy\\MCC}} & \multirow{4}{*}{zh} & StcokA  &  1,477  &\multirow{7}{*}{$\times$}&\multirow{7}{*}{$\times$}&\multirow{7}{*}{$\times$ }&\multirow{7}{*}{$\times$ } &\multirow{7}{*}{$\times$} &\multirow{7}{*}{$\times$} & \multirow{7}{*}{$\checkmark$}&\multirow{7}{*}{$\times$}&\multirow{7}{*}{$\times$}& \multirow{7}{*}{$\checkmark$}   \\
                                         &            &                       &                      &CACL18      &  511                \\
                                         &            &                        &                     & CBigData18 &  159                   \\
                                         &            &                        &                     & CCIKM18    &  86                    \\
\cmidrule(lr){4-6}
                                         &            &                        & \multirow{3}{*}{en} & ACL18      &  3,720                    \\
                                         &            &                        &                     & BigData18  &  1,472                    \\
                                         &            &                        &                     & CIKM18     &  1,143                     \\
\midrule
\makecell{Relationship\\Extraction}   &\rotatebox[origin=c]{90}{FinRE}   & \makecell{F1\\Accuracy}   & zh  & RE  &  1,489   &$\times$   &$\checkmark$ &$\times$&$\times$  &$\times$ &$\times$  &$\times$&$\times$&$\times$ &$\times$ \\
\midrule
\makecell{Headline\\Classification} &\rotatebox[origin=c]{90}{FinHC} &  Avg F1 & \makecell{zh\\en}  & \makecell{CHeadlines\\Headlines} &  \makecell{2,051\\20,547}& $\times$ &$\times$&$\times$&$\times$ &$\times$  & $\checkmark$ &$\checkmark$ &$\times$ &$\times$& $\checkmark$ \\
\midrule
\makecell{Credit\\Classification} &\rotatebox[origin=c]{90}{FinCC}   &\makecell{F1\\Accuracy\\MCC} & en & \makecell{German\\Australian}  & \makecell{200\\139} & $\times$ & $\times$& $\times$&$\times$& $\times$ &$\times$& $\checkmark$ &$\times$ &$\times$& $\checkmark$ \\
\midrule  
\makecell{Hawkish-dovish\\Classification}   & \rotatebox[origin=c]{90}{FinDC}  &\makecell{F1\\Accuracy}  & en  & FOMC  &  496    &$\times$ &$\times$&$\times$ &$\times$  &$\times$ &$\times$& $\checkmark$ &$\times$&$\times$& $\checkmark$   \\
\midrule
\multirow{5}{*}{\makecell{Question \\Answering}}      & \multirow{5}{*}{\rotatebox[origin=c]{90}{FinQA}} &\multirow{5}{*}{EM Accuracy}  & \multirow{3}{*}{zh} & QA &  2,469 &\multirow{5}{*}{$\checkmark$}&\multirow{5}{*}{$\checkmark$}&\multirow{5}{*}{$\times$} &\multirow{5}{*}{$\checkmark$} &\multirow{5}{*}{$\times$}&\multirow{5}{*}{$\checkmark$}& \multirow{5}{*}{$\checkmark$}&\multirow{5}{*}{$\checkmark$}&\multirow{5}{*}{$\checkmark$}&\multirow{5}{*}{$\checkmark$}  \\
                                         &              &                      &                     & CEnQA     &  133                   \\
                                         &              &                      &                     & CConFinQA  &  237                    \\
\cmidrule(lr){4-6}
                                         &             &                       & \multirow{2}{*}{en} & EnQA      &  1,147                    \\
                                         &             &                       &                     & ConFinQA   &  1,490                    \\
\midrule
\makecell{Entity \\Recognition}     & \rotatebox[origin=c]{90}{FinER} &\makecell{F1\\Entity F1}& \makecell{zh\\en\\en} & \makecell{CNER\\NER \\FINER-ORD} &\makecell{337\\98\\1,075}  &$\checkmark$&$\checkmark$&$\times$&$\times$&$\times$&$\checkmark$& $\checkmark$ &$\times$&$\times$&$\checkmark$  \\
\midrule
\multirow{4}{*}{\makecell{Event \\Detection}}         & \multirow{4}{*}{\rotatebox[origin=c]{90}{FinED}} &\multirow{4}{*}{\makecell{F1\\Recall\\Precision}} & \multirow{4}{*}{zh} & 19CCKS & 2,936 &\multirow{4}{*}{$\checkmark$}&\multirow{4}{*}{$\times$}&\multirow{4}{*}{$\times$}&\multirow{4}{*}{$\times$}&\multirow{4}{*}{$\times$} &\multirow{4}{*}{$\times$}& \multirow{4}{*}{$\times$}&\multirow{4}{*}{$\times$} &\multirow{4}{*}{$\times$} &\multirow{4}{*}{$\times$}                \\
                                         &              &                      &                     & 20CCKS     &  9,159                   \\
                                         &              &                      &                     & 21CCKS     &  1,400                   \\
                                         &              &                      &                     & 22CCKS     &  11,829                    \\
\midrule
\makecell{Text \\Summarization}      &\rotatebox[origin=c]{90}{FinTS} &\makecell{Rouge\\BERTScore\\BARTScore}  & \makecell{zh\\en\\en} & \makecell{NA\\ECTSUM\\EDTSUM}&\makecell{3,600 \\495\\2000}  &$\checkmark$&$\checkmark$&$\times$&$\times$&$\times$ &$\times$  &$\checkmark$  &$\times$&$\times$ &$\checkmark$              \\ 
\midrule
\multicolumn{6}{c|}{New Task Total}   & 0& 2& 0 &0 &0& 1& 0&0 &1 & 1     \\
\midrule
\multicolumn{2}{c}{Open Source}  &\multicolumn{4}{|c|}{$\checkmark$}  & $\checkmark$& $\checkmark$ &$\checkmark$ &$\times$ & $\checkmark$& $\checkmark$& $\checkmark$&$\times$&$\times$& $\checkmark$   \\
\multicolumn{2}{c}{Task Total}  &\multicolumn{4}{|c|}{14}    & 6& 8& 1 &1 &1& 4& 8&1 &2 & 8 \\
\multicolumn{2}{c}{Data Total}  &\multicolumn{4}{|c|}{95k} &4k& 11k&5k &0.2k &9.9k&4k& 43k&10k &5.5k&54k  \\
\bottomrule[1pt]
    \end{tabular}
    \label{tab:eva}
\end{table}

Following previous methods~\cite{zhang2024d,xie2023pixiu}, we integrate multiple metrics, such as weighted F1 Score (F1), accuracy (ACC), entity-level F1 score (Entity F1), EM (Exact Match) Accuracy, and the Matthews correlation coefficient (MCC), to evaluate performance.

\section{Experiments}

In addition to the FinLLMs  such as CFGPT-sft (CFGPT), DISC-FinLLM (DISCFIN), FinMA, we also select the most representative baseline LLMs for performance evaluation on ICE-FLARE.

\begin{enumerate}
\item  \textbf{GPT-4}~\cite{achiam2023gpt}: An OpenAI-developed powerful LLM with around 1T parameters, capable of handling diverse modalities including images and videos.
\item  \textbf{ChatGPT}~\cite{brown2020language}: It also known as ChatGPT-3.5-Turbo-1106 with 175B parameters from OpenAI, is a widely-known LLM in the NLP domain.
\item  \textbf{LLaMa}~\cite{touvron2023llama}: LLaMa and advanced LlaMa2 are powerful LLMs developed by Meta AI, with parameter sizes ranging from 7 billion to 650 billion.
\item  \textbf{Baichuan}~\cite{yang2023baichuan}: This LLM developed by Baichuan-inc for Chinese and English NLP tasks, available in Baichuan-7B and Baichuan-7B-Chat variants.
\item  \textbf{ChatGLM} ~\cite{du2021glm}: It is an advanced bilingual LLM developed by Tsinghua University and ZhiPu AI, with variants like ChatGLM2-6B and ChatGLM3-6B.
\item  \textbf{BLOOM} ~\cite{muennighoff2022crosslingual}: Developed by the BigScience team, it is a versatile LLM with parameter sizes ranging from 560M to 176B, including  Bloomz-7B1.
\item  \textbf{InternLM}~\cite{team2023internlm}: InternLM is a language model developed by SenseTime, featuring different parameter configurations such as InternLM-7B.
\item  \textbf{Qwen}~\cite{bai2023qwen}: It is a set of advanced LLMs introduced by Alibaba Cloud, including Qwen-7B for conversational capabilities in Chinese. 
\end{enumerate}

All results from all LLMs are conducted in the zero-shot setting~\cite{li2023chatgpt} with automatic evaluation method. 

\begin{table*}[htb!]
\centering
\footnotesize
\setlength{\tabcolsep}{1.3pt}
\renewcommand{\arraystretch}{0.8}
\caption{The zero-shot performance of different 6B-7B LLMs, baseline FinLLMs and variants of ICE-INTERN on the ICE-FLARE benchmark. All results are based on the average values of the NLP tasks including classification (CLS), extraction (EXT), reasoning (REA), prediction (PRE), generation (GEN),  translation (TRA), and  data-out-of-training (DOT) task serving as the final evaluation criteria of ICE-FLARE benchmark. Results with "\_\_" indicate the best results among models excluding ChatGPT.}
   \vspace{-5pt}
    \begin{tabular}{l|ccccc|c|ccccc|c|c|c|c|c}
    \toprule[1pt]
    Model              & ZH-CLS   & ZH-PRE   & ZH-EXT   & ZH-GEN    & ZH-TRA   & ZH.Avg & EN-CLS   & EN-PRE   & EN-EXT   & EN-REA   & EN-OFT    & EN.Avg & Bilingual.Avg & CLS.Avg & PRE.Avg & EXT.Avg   \\
    \hline
    Bloomz-7B1         & 0.170 & 0.131 & 0.038 & -1.650 & 0.240 & -0.214 & 0.406 & 0.238 & 0.000 & \underline{0.005} & -1.238 & -0.118 & -0.166 &0.288&0.185&0.019\\
    Baichuan-7B        & 0.202 & 0.000 & 0.005 & -2.097 & 0.241 & -0.330 & 0.264 & 0.248 & 0.000 & 0.000 & -1.246 & -0.147 & -0.238 &0.233&0.124&0.003\\
    Baichuan2-7B       & 0.005 & 0.000 & 0.001 & -2.262 & 0.179 & -0.415 & 0.173 & 0.226 & 0.000 & 0.000 & -1.307 & -0.182 & -0.299 &0.089&0.113&0.001\\
    LLaMa2-7B          & 0.205 & 0.000 & 0.001 & -1.262 & 0.294 & -0.152 & 0.271 & 0.243 & 0.012 & 0.000 & -1.214 & -0.138 & -0.145 &0.238&0.122&0.007\\
    ChatGLM2-6B        & 0.036 & 0.000 & 0.005 & -2.265 & 0.179 & -0.409 & 0.205 & 0.225 & 0.000 & 0.000 & -1.298 & -0.174 & -0.291 &0.121&0.113&0.003\\
    ChatGLM3-6B        & 0.391 & 0.220 & 0.013 & -1.031 & 0.311 & -0.019 & 0.434 & 0.262 & 0.248 & 0.000 & \underline{-0.833} & \underline{0.022}  & 0.002  &0.413&0.241&0.131\\
    Qwen-7B            & 0.327 & 0.154 & 0.001 & -1.185 & 0.295 & -0.082 & 0.454 & 0.226 & 0.012 & 0.000 & -1.173 & -0.096 & -0.089 &0.391&0.190&0.007\\
    ChatGPT            & 0.432 & 0.422 & 0.039 & -1.693 & 0.311 & -0.098 & 0.430 & 0.262 & 0.770 & 0.590 & -0.719 & 0.267  & 0.084  &0.431&0.342&0.405\\
    GPT-4              & 0.549 & 0.383 & 0.098 & -1.577 & 0.266 & -0.056 & \textbf{0.618} & \textbf{0.283} & \textbf{0.830} & \textbf{0.695} & \textbf{-0.640} & \textbf{0.357}  & \textbf{0.151}  &\textbf{0.584}&0.333&0.464\\
    InternLM-7B        & 0.277 & 0.003 & 0.001 & -1.119 & 0.272 & -0.113 & 0.301 & 0.236 & 0.000 & 0.000 & -1.251 & -0.143 & -0.128 &0.289&0.120&0.001\\
    CFGPT-7B & 0.211 & 0.050 & 0.011 & -2.017 & 0.272 & -0.295 & 0.242 & 0.224 & 0.000 & 0.000 & -1.265 & -0.160 & -0.227 &0.227&0.137&0.006\\
    DISCFin-13B        & 0.290 & 0.315 & 0.022 & -1.858 & 0.290 & -0.188 & 0.446 & 0.236 & 0.039 & 0.002 & -1.158 & -0.087 & -0.138 &0.368&0.276&0.031\\
    FinMA-7B      & 0.229 & 0.000 & 0.016 & -2.009 & 0.250 & -0.303 & 0.371 & 0.233 & \underline{0.392} & 0.000 & -1.187 & -0.038 & -0.171 &0.300&0.117&0.204\\
    \hline
    ICEdlc-7B  & 0.598 & 0.340 & 0.020 & -1.762 & 0.337 & -0.093 & 0.331 & \underline{0.282} & 0.000 & 0.000 & -1.227 & -0.123 & -0.108 &0.465&0.311&0.010\\
    ICEdle-7B  & 0.607 & \textbf{0.637} & \textbf{0.569} & -1.024 & 0.327 & 0.223  & 0.334 & 0.275 & 0.006 & 0.000 & -1.203 & -0.118 & 0.053  &0.471&\textbf{0.456}&\underline{0.288}\\
    ICEdtt-7B  & \underline{0.607} & \underline{0.630} & 0.566 & \underline{-1.024} & \textbf{0.448} & \underline{0.245}  & 0.357 & 0.268 & 0.000 & 0.000 & -1.182 & -0.111 & 0.067  &0.482&\underline{0.449}&0.283\\
    ICEfull-7B & \textbf{0.631} & 0.617 & \underline{0.567} & \textbf{-0.972} & \underline{0.409} & \textbf{0.250}  & \underline{0.494} & 0.250 & 0.362 & 0.000 & -1.193 & -0.017 & \underline{0.117} &\underline{0.563}&0.434&\textbf{0.465}\\
    \bottomrule[1pt]
    \end{tabular}
    \label{tab:ben}
\end{table*}

\begin{figure*}[htb!]
    \centering
    \includegraphics[width=7in]{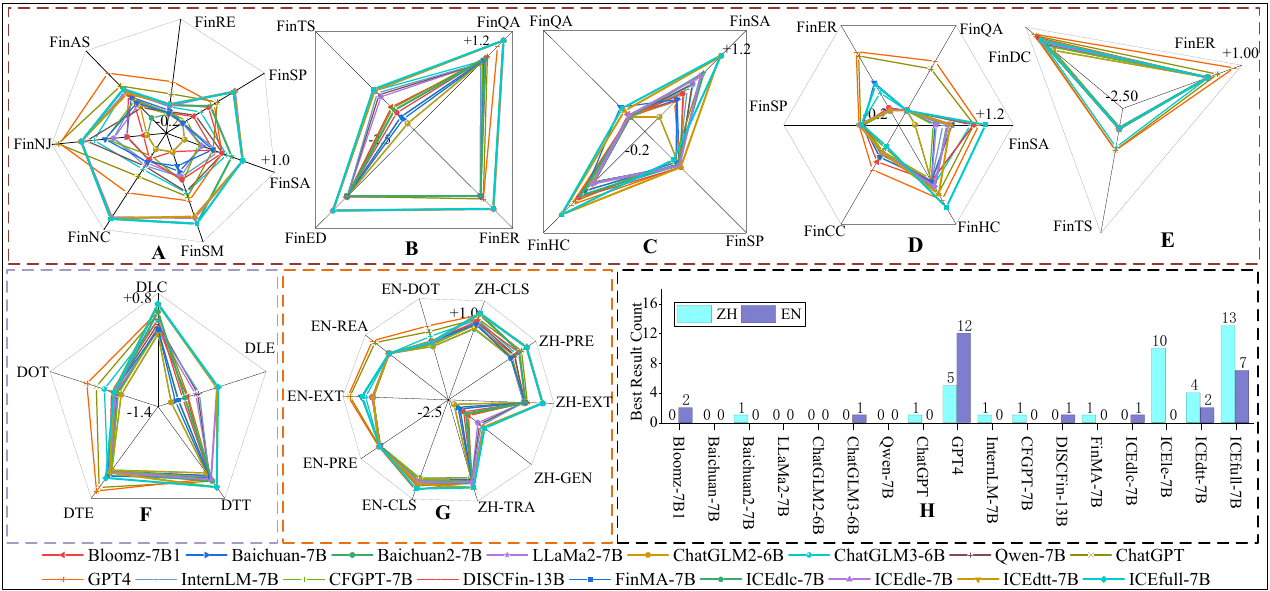}
    \vspace{-5pt}
    \caption{Seven radar charts respectively demonstrate the average metrics on the specific tasks contained within different data types (DLC (A), DLE (B), DTT (C), DTE (D), DOT (E)), the average metrics across 5 data types (F), and the average metrics on 10 bilingual NLP tasks (G), along with a histogram showing the count of best results on Chinese (ZH) and English (EN) data (H).}
    \label{fig:res}
\end{figure*}

\subsection{Results}
\label{sec:results}

The results for ICE-FLARE benchmark  are shown in Table \ref{tab:ben} and Figure \ref{fig:res}. More detailed results can be found in Appendix Table \ref{tab:res}.

\subsubsection{\textbf{Overall Performance}} 
As shown in Table \ref{tab:ben}, ICE-INTERN variants outperform all baseline LLMs in Chinese tasks, even surpassing larger LLMs like GPT-4. ICE-INTERN falls slightly behind GPT-4 in average performance across bilingual tasks but surpasses all other LLMs. It excels in CLS and PRE bilingual tasks, outperforming all models. However, its performance in REA and OFT tasks is not particularly remarkable, sometimes even weaker than other LLMs such as FinMA-7B and Bloomz-7B1. This clearly demonstrates the absolute superiority of ICEfull-7B in the Chinese-English financial bilingual domain, emphasizing the importance of instruction fine-tuning over the InternLM backbone model in optimizing the financial performance of various LLMs. ICE-INTERN variants like ICEdle-7B and ICEdtt-7B, which are not fine-tuned with English datasets, thus performing slightly below GPT-4 due to its larger parameters and emphasis on English. This suggests that there is further optimization potential in future model iterations by exploring the introduction of more effective English data. Referring to Table \ref{tab:res}, ICE-INTERN  variants achieve the best performance in 68.25\% (excluding ChatGPT and GPT-4) and 58.73\% of the 63 specific-task metrics. In terms of overall best results, the fully fine-tuned ICEfull-7B outperforms all baseline models, notably outperforming both ChatGPT and even surpassing larger models like GPT-4.
 
\subsubsection{\textbf{Language Disparity}}  
Table \ref{tab:ben} highlights significant language discrepancies in how current LLMs, such as well-known ChatGPT, tackle Chinese-English bilingual financial NLP tasks. The findings suggest that these LLMs, including the Chinese-focused Qwen-7B, excel predominantly in English but exhibit slight weaknesses in Chinese. Our proposed ICE-INTERN and its variations distinctly prioritize Chinese proficiency over English, both in individual tasks and across overall average metrics. This indicates that addressing language disparities in LLMs requires dedicated efforts to construct and optimize high-quality instructions with multilingual characteristics. Further referencing Figure \ref{fig:res}, \textbf{A}-\textbf{C} indicates that ICEfull-7B excels in specific tasks involving DLC, DLE, and DTT data types, showcasing strong Chinese language capabilities. However, in \textbf{D}-\textbf{E}, it slightly lags behind GPT-4 in the FinER, FinQA, FinTS, FinCC tasks, with the performance gap being particularly pronounced in \textbf{F}. In \textbf{G}, ICEfull-7B excels in 10 Chinese-English bilingual NLP tasks, except for English extraction (EN-EXT), reasoning (EN-REA) and data-out-of-training (EN-DOF) tasks. \textbf{H} further confirms ICEfull-7B's overall superiority in handling tasks in both English and Chinese, with particular excellence in Chinese. 

\subsubsection{\textbf{Generalization Ability}} 
Despite the strengths exhibited by ICE-INTERN and its variants on known specific tasks, they lags behind GPT-4 in unknown tasks like FinDC, FINER, and FinST within the unfine-tuned DOF data type. This performance gap indicates the need for a wider variety of domain-specific financial tasks for effective model fine-tuning. Specifically,  we employ complex prompt designs involving directly generating label sequences in the FinST task. However, our fine-tuned ICE-INTERN models consistently encounter difficulties in producing outputs in the desired formats. Even LLMs such as larger-parameter ChatGPT and GPT-4, which demonstrate some capability for token labeling in FinER, encounter challenges in generating labels and summaries for both the ECTSUM and EDTSUM datasets in FinST, especially when dealing with longer financial contextual information. Our findings are in line with the conclusions put forth in existing studies~\cite{xie2023pixiu}, suggesting that for unfamiliar financial tasks within current FinLLMs, it will be imperative to provide more extensive domain-specific tasks and instruction data in the future to further enhance model performance.

\subsubsection{\textbf{Ablation study}}  
This ablation study primarily analyzes the performance impact of variant models on different datasets (DLC, DLE, DTT, DTE, and DOT). As shown in Table \ref{tab:res}, DLC has minimal impact on the performance of ICEdlc-7B, while fine-tuning with DLE significantly improves the performance of ICEdle-7B. Conversely, fine-tuning ICEdle-7B with DTT does not enhance the performance of ICEdtt-7B and may even lead to a decrease. Finally, full fine-tuning of ICEFull-7B with DTE shows a significant performance improvement. This indicates that InternLM, as the backbone model, is already capable of handling most simple financial classification tasks present in DLC data types, making fine-tuning on such data less consequential. Clearly, it is essential to incorporate a variety of DLE data types in the instruction data to enhance model performance for comprehension and inference in more intricate tasks. Furthermore, we find that in the cross-lingual financial classification tasks in FinHC and FinSA, the ICEdtt-7B model shows performance improvements of approximately 1\% to 10\% compared to ICEdle-7B when incorporating Chinese translation datasets. However, for the complex FinSP task, the performance exhibits a mix of improvements and declines. This is because tasks like FinHC and FinSA involve relatively simple financial description syntax and structure, allowing ICEdtt-7B to integrate translated data to learn sufficient information for answer reasoning. However, for tasks like FinSP, which entail longer and more complex financial text including historical stock prices and tweets, data augmentation methods did not yield the anticipated model performance enhancements. This emphasizes the need to consider text length and task complexity when deciding on incorporating translation data during LLM fine-tuning.

\subsection{Practical Examples}
\label{sec:exmaples}

To demonstrate the practical utility of ICE-PIXIU, we provide examples of model responses of ICE-PIXIU and other baseline including GPT-4 and Baichuan-7B, as outlined in in Appendix Table \ref{tab:llms}. It can be concluded that ICE-INTERN provides more accurate responses in FinED, FinNC, and FinRE tasks, and offers more comprehensive responses in FinER and FinQA tasks. Particularly in FinER, ICE-INTERN is capable of providing all entity names and their corresponding types. Overall, ICE-INTERN demonstrates superior in accuracy and comprehensiveness in financial question answering.

\section{Limitations}
\label{sec:limitations}

While this study has positive aspects, we acknowledge limitations:
1) \textbf{Parameter Constraints}: The limited resources hinder training models with over 7B parameters, impacting performance on large datasets.
2) \textbf{Backbone LLM Adaptability}: The fine-tuned model performance varies based on the backbone model adaptability to specific tasks, requiring task-specific selection.
3) \textbf{Prompt Quality Variability}: Even with expert annotation, the quality of instruction prompts may perform vary across models, affecting model assessment.
4) \textbf{Risk of Misuse}: Open-source models facilitate research, but commercial misuse raises concerns about financial risks. Effective regulation is therefore necessary to mitigate these risks.

\section{Conclusion}
\label{sec:conclusion}
We introduce a Chinese-English bilingual financial framework ICE-PIXIU featuring ICE-INTERN LLM, ICE-FIND instruction data, and ICE-FLARE evaluation benchmark to address global financial cross-lingual disparities. ICE-INTERN's bilingual capability empowers global finance by breaking language barriers, while ICE-FLARE enables comprehensive cross-lingual assessments for various financial tasks. Notably, ICE-INTERN demonstrates superior performance over existing SOTA LLMs, including GPT-4, in Chinese financial tasks by effectively leveraging strategic instruction tuning and a diverse dataset for cross-linguistic transfer. ICE-PIXIU underscores the significance of diverse instruction data types and cross-lingual data translation transfer in enhancing performance. By rigorously evaluating DOT datasets, ICE-PIXIU supports robustness exploration of FinLLMs and ensures consistent performance assessments. ICE-PIXIU's open-access framework drives exploration in financial NLP for complex tasks and multilingual applications. 



\bibliographystyle{ACM-Reference-Format}
\bibliography{acmart}

\section*{Appendix}

\subsection*{A. A Example of Translating BigData22 to English}
\label{sec:appendixfigtra}

\begin{figure*}[htb!]
	\centering 
	\includegraphics[width=5.5in]{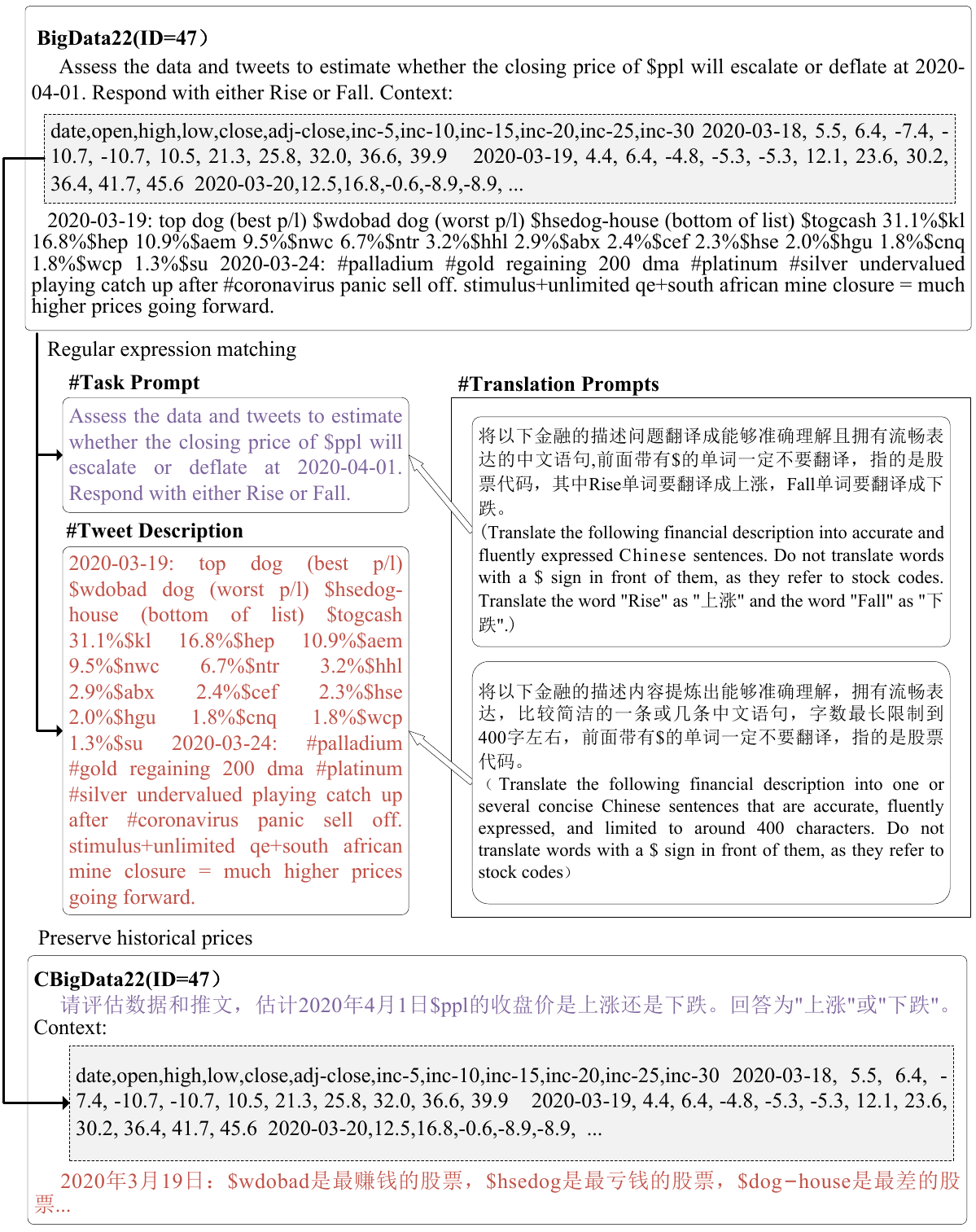}
	\caption{Giving the example of translating BigData22 English dataset into CBigData22 Chinese dataset with human-written specific translation prompts. Specifically, we will translate the task prompts and twitter descriptions from BigData22 separately, while preserving the original stock information, which includes stock tickers with \$ signs and historical stock prices, unchanged.}
   \label{fig:tra}
\end{figure*}

Figure \ref{fig:tra} presents the examples of our strategy of translating the BigData22 dataset into Chinese using human-written specific prompts

\subsection*{B.  Examples of Instruction Prompts in Chinese Task}
\label{sec:appendixitd}

Table \ref{tab:tra} displays examples of Chinese human-written prompts and their corresponding English translations for each specific task.

\begin{table*}[htb!]
    \centering
    \footnotesize
    \setlength\tabcolsep{4pt}
    \renewcommand{\arraystretch}{3}
    \caption{Examples of the Chinese prompt and corresponding English translation for various specific task in Chinese financial datasets.}
    \begin{tabular}{m{2.8cm}<{\centering}|m{6cm}|m{8cm}}
    \toprule[1pt]
    \textbf{Specific Task}       & \multicolumn{1}{c|}{\textbf{Chinese Prompt}}    &  \multicolumn{1}{c}{\textbf{English Translation}}    \\ 
    \midrule
    \makecell{Sentiment Analysis \\(FinSA)}    & 确定所提供的金融新闻文章中的句子的情绪，识别其中的情绪是积极，消极，还是中性的，你的回答应该是其中一个。  & Determine the mood of the sentences in the financial news articles provided and identify whether the mood is 'Positive', 'Negative' or 'Neutral'. Your answer should be one of them.	     \\
    \midrule
    \makecell{Sentiment Matching \\(FinSM)}    & 判断给出的两个金融文本表达的意思是否相似，你只需要回答是或否。 &  To judge whether the two financial texts are similar in meaning, you only need to answer 'Yes' or 'No'.  \\
    \midrule
    \makecell{News Classification \\(FinNC)}    & 根据金融报道的内容，判断其属于'中国'、'国际'还是'外国'，只需给出其中一个类别。现在分析下面的报道，回答它所属的类别： & According to the content of the financial report, determine whether it belongs to 'China', 'International', or 'Foreign', only one category needs to be given. Now analyze the following report and answer the category it belongs to:   \\
    \midrule
    \makecell{Negative Judgment \\(FinNJ)}    & 根据所给的金融新闻及实体，你需要判断所给实体是否含有负面的消息，你只需简单回答'有'或者'无'   &  Based on the given financial news and entities, you need to determine whether the given entities contain negative messages. You only need to answer 'Yes' or 'No'.    \\
    \midrule
    \makecell{Answer Selection \\(FinAS)}      &  根据所提出的金融问题，从以下四个选项中选择最合适的一个。你的输出应该是：'A'、'B'、'C' 或'D'。 &  Based on the financial issue presented, choose the most appropriate one from the following four options. Your output should be: 'A', 'B', 'C', or 'D'.    \\
    \midrule
    \makecell{Stock Prediction \\(FinSP)}        & 请仔细分析数据和推文，预测2017-10-13时\$trv的收盘价格是上涨还是下跌。请确认是上涨还是下跌, 只回答为'上涨'或者'下跌'。 &  Please carefully analyze the data and tweets, and predict whether the closing price of \$TRV on 2017-10-13 will go up or down. Please confirm whether it is going up or down. You can answer 'Rise' or 'Fall'.       \\
    \midrule
    \makecell{Headline Classification \\(FinHC)} &  请考虑标题是否讨论了与黄金相关的过去事件。新闻标题是否暗示黄金的过去新闻？您的回答应该为'是'或'否'。   & Please consider whether the headline discusses past events related to gold. Does the news headline imply past news about gold? Your answer should be 'Yes' or 'No'.	     \\
    \midrule
    \makecell{Question Answering \\(FinQA)}      &  你需要分析金融文本，根据内容回答相关问题。如果没有对应的答案，可以回答'无相应参数'。   & You need to analyze the financial text and answer relevant questions according to the content. If there is no corresponding answer, you can answer 'No corresponding parameter'.    \\
    \midrule
    \makecell{Event Detection\\ (FinED)}      & 阅读以下金融领域的公告，判断所有的事件类型及其对应主体，省略数字。请以:'事件类型, 事件主体'的格式回复 。其中事件类型应该在这里面：['信批违规',...,'涉嫌欺诈']   &   The task is to read a financial announcement, determine all event types and their corresponding entities, and reply in the format 'event type, event subject'. The event types should be among those listed in the brackets [‘credit approval violation’,...,’suspected fraud’].  \\
    \midrule
    \makecell{Entity Recognition \\(FinER)}     &  在分析中国证券交易委员会备案文件中的句子时，识别指明个人('PER')、组织('ORG')或地点('LOC')的特定命名实体。答案应遵循格式'实体名称, 实体类型' 。  &   When analyzing sentences in the China Securities Regulatory Commission filing documents, identify specific named entities that indicate individuals ('PER'), organizations ('ORG'), or locations ('LOC'). The answer should follow the format 'entity name, entity type'.  \\
    \midrule
    \makecell{Relationship Extraction \\(FinRE)} &  请仔细分析所给金融报道和实体对，然后在['合并', '竞争',...]中选择能准确描述该实体对关系的选项。请直接给出答案，如有疑虑可回答unknown。  &   Please carefully analyze the given financial report and entity pair, then choose the option from ['merge', 'compete', ...] that accurately describes the relationship of the entity pair. Please provide the answer directly; if in doubt, you may respond with 'unknown'                     \\
    \midrule
    \makecell{Text Summarization \\(FinTS)}      &  通过阅读金融公告，你的任务是对所给的文本进行简短的总结，重点突出主要论点，长度保持在一到两句之间。 &  Your task is to provide a brief summary of the given text by reading financial announcements, emphasizing the main arguments, and keeping the length between one to two sentences  \\  
    \bottomrule[1pt]
    \end{tabular}
\label{tab:tra}
\end{table*}

\subsection*{C. Examples of Introduction for Each Financial Task}
\label{sec:appendixexa}

Table \ref{tab:exa1} to Table \ref{tab:exa3} showcase examples of contextual text descriptions and target answers in various financial application scenarios.

\begin{table*}[htb!]
    \centering
      \footnotesize
     \setlength\tabcolsep{3pt}
    \renewcommand{\arraystretch}{1.5}
    \caption{Examples with contexts and target answers in Sentiment Analysis(FinSA), Sentiment Matching(FinSM), News Classification(FinNC), Negative Judgment(FinNJ), Headline Classification(FinHC), Answer Selection(FinAS) and Question Answering(FinQA).}
    \begin{tabular}{m{2cm}<{\centering}|m{15cm}}
    \hline
    FE & Context: 国检集团和华测检测哪个比较漂亮？业务都差不多，都是检测，并且国检持有碳交易股份，两个都买了。  \\ 
                          & Answer: 金融的报道偏向\textcolor{blue}{积极}评价。 \\ 
    \hline
    StockB & Context: 中烟集团金融网站,融通租赁注册100亿！揭牌仪式有望。  \\ 
                          & Answer: 金融的报道偏向\textcolor{blue}{中性}评价。 \\
    \hline
    CFPB &  Context: 维萨拉在2007年第三季度的净利润从2006年同期的680万欧元（980万美元）降至300万欧元（430万美元）。  \\ 
                          &  Answer: 金融的报道偏向\textcolor{blue}{消极}评价。 \\ 
    \hline
    CFiQA-SA & Context: \$CERN-在50和200MA上方整合,这里是很好的长期进入点，止损位于10MA以下-目标区域\$70。  \\ 
                          & Answer: 金融的报道偏向\textcolor{blue}{积极}评价。 \\ 
    \hline
    FPB & Context: Hearst will be able to consolidate about 20\% of all Russian market for advertising in press after the purchase. \\ 
                          & Answer: Sentiment analysis of financial news is \textcolor{blue}{positive}. \\ 
    \hline
    FiQA-SA & Context: \$SLV-4.44\% at 18 now AWFUL, down from 42.50  \\ 
                          & Answer: Sentiment analysis of financial news is \textcolor{blue}{negative}. \\ 
    \hline
    BQC& Context: 1:两个小时还没有等到确认电话怎么办？明天会继续联系嘛？ 2:下次借款是否不需要电话确认   \\ 
                          & Answer: 两个金融表达语义\textcolor{blue}{不是}相似的。 \\ 
    \hline
    AFQMC & Context: 1:借呗还款日当天不能再借款吗 2:蚂蚁借呗要一次性还清才能再借吗   \\ 
                          & Answer: 两个金融表达语义\textcolor{blue}{是}相似的。 \\ 
    \hline
    NL & Context: [中航泰达：拟购买包钢节能34\%股权]中航泰达公告，公司拟以2.09亿元现金认购包钢节能新增注册资本1.28亿元，同时以2.59亿元受让北方稀土持有标的公司的1.59亿元注册资本。本次交易前，北方稀土直接持有包钢节能100\%股权。本次交易完成后，公司将直接持有包钢节能34\%的股权。   \\ 
                          & Answer: 报道涉及\textcolor{blue}{中国}区域。 \\ 
    \hline
    NL2 & Context: 欧股集体高开，德国DAX30指数涨0.44\%，英国富时100指数涨1.29\%，法国CAC40指数涨0.7\%，欧洲斯托克50指数涨0.63\%。   \\ 
                          & Answer: 报道涉及\textcolor{blue}{国际大盘}。 \\ 
    \hline
    NSP & Context: 钱宝网张小雷涉嫌集资诈骗罪被检方提起公诉中国基金报。实体：钱宝网;钱宝   \\ 
                          & Answer: 所给实体\textcolor{blue}{是}含有负面消息。 \\ 
    \hline
    CHeadlines & Context: 在亚洲早盘，金价略有下跌，市场正在关注美国的数据。   \\ 
                          & Answer: 标题中\textcolor{blue}{是}提到金价会下降的意见。 \\ 
    \hline
    Headlines & Context: Gold holds near 3-1/2 week low as investors opt for riskier assets.  \\ 
                          & Answer: \textcolor{blue}{Yes}, the headline suggests a downward direction for gold. \\ 
    \hline
    FinevalF & Context: 国债的偿债率是指()?   \\ 
                          & Answer: 答案是\textcolor{blue}{C}:当年还本付息额占当年财政收入的比重。 \\ 
    \hline
    QA  & Context: 11月9日上午，佛山市人民政府与徐工集团工程机械有限公司签约，将在南海区建设总投资20亿元的广东生产基地项目。   \\ 
                          & Answer: 建设投资方是\textcolor{blue}{徐工集团}。   \\ 
    \hline
    CEnQA & Context: 下表比较了花旗五年普通股的累积总回报，该股在纽约证券交易所上市，代码为201cc201d。日期，花旗，标普500，标普金融指数；2012-12-31，100.0，100.0，100.0; 2013-12-31，131.8，132.4，135.6；2014-12-31，137.0，150.5，156.2；2015-12-31，131.4，152.6，153.9；2016-12-31，152.3，170.8，188.9；2017-12-31，193.5，208.1，230.9。  \\ 
                          & Answer: 五年回报百分比\textcolor{blue}{0.935}。   \\ 
    \hline
    CConFinQA & Context: 以下是比较2007年和2008年净收入变动的分析，金额（以百万计）。<table class='wikitable'><tr><td>1</td><td></td><td>金额（以百万计）</td> </tr> <tr> <td>2</td ><td>2007年净收入</td> <td>\$991.1</td> </tr> <tr><td>3</td> <td>零售电价</td> <td>-17.1（17.1）</td> </tr> <tr> <td>4</td> <td>购买的电力容量</td> <td>-12.0（12.0）</td> </tr> <tr> <td>5</td> <td>净批发收入</td> <td>-7.4（7.4）</td> </tr> <tr> <td>6</td> <td>其他</td> <td>4.6</td> </tr> <tr> <td>7</td><td>2008年净收入</td> <td>\$959.2</td> </tr> </table>。   \\ 
                          & Answer: 2007年的净收入是\textcolor{blue}{\$991.1}。   \\ 
    \hline
    EnQA & Context: The following table provides a comparison of the accumulated total return of citi common stock over a period of five years, which is listed on the nyse under the ticker symbol 201cc201d. date, citi, s\&p 500, s\&p financials. 31-dec-2011, 100.0, 100.0, 100.0; 31-dec-2012, 150.6, 116.0, 128.8; 31-dec-2013, 198.5, 153.6, 174.7; 31-dec-2014，206.3, 174.6, 201.3; 31-dec-2015, 197.8, 177.0, 198.2; 31-dec-2016, 229.3, 198.2, 243.4.  \\ 
                          & Answer: The percent of the growth for s\&p financials cumulative total return from 2013 to 2014 is \textcolor{blue}{26.6}.  \\ 
    \hline
    ConFinQA & Context: The following is an analysis comparing the changes in net income between 2007 and 2008. <table class='wikitable'><tr> <td>1</td> <td> </td> <td> amount (in millions)</td> </tr> <tr><td>2</td> <td>2007 net revenue</td> <td>\$991.1</td> </tr> <tr> <td>3</td><td>retail electric price</td> <td>-17.1 ( 17.1 )</td> </tr> <tr> <td>4</td> <td>purchased power capacity</td> <td>-12.0 ( 12.0 )</td> </tr> <tr> <td>5</td> <td>net wholesale revenue</td><td>-7.4 (7.4)</td> </tr> <tr> <td>6</td> <td>other</td> <td>4.6</td></tr> <tr> <td>7</td> <td>2008 net revenue</td> <td>\$959.2</td> </tr> </table>。   \\ 
                          & Answer: The net revenue in 2008 is \textcolor{blue}{\$959.2}. \\ 
    \hline
    \end{tabular}
     \label{tab:exa1}
\end{table*}

\begin{table*}[htb!]
    \centering
      \footnotesize
     \setlength\tabcolsep{3pt}
    \renewcommand{\arraystretch}{3}
    \caption{Examples with contexts and target answers in  Stock Prediction(FinSP).}
    \begin{tabular}{m{2cm}<{\centering}|m{15cm}}
    \hline
    StockA & Context: 新北洋公告，公司控股子公司荣鑫科技董事会审议通过了《关于拟申请公司股票在全国中小企业股份转让系统终止挂牌的议案》，具体详见荣鑫科技（证券代码：839288）披露在全国中小企业股份转让系统（以下简称“新三板”）的相关公告。荣鑫科技基于其自身经营发展及战略规划的需要，拟申请在新三板终止挂牌。 日期 开盘价 闭盘价 2021-01-04 34.63 34.9763 2021-01-05 34.63 35.5122 2021-01-06 35.1659 34.0116 2021-01-07 33.8879 32.3048 2021-01-08 32.0987 32.3213.  \\ 
                          & Answer: 该股票\textcolor{blue}{表现不佳}。 \\ 
    \hline
    CACL18 & Context: \$codi近期的走向趋势如下，open, high, low, close, adj-close, inc-5, inc-10, inc-15, inc-20, inc-25, inc-30；2015-09-16, -0.8, 0.2, -1.3, 1.0, 1.0, -0.9, -0.1, -0.1, -0.3, 0.0, 0.5; 2015-09-17,-1.4, 0.9, -1.7, 1.6, 1.6, -2.0, -1.7, -1.5, -1.8, -1.6, -1.2; 2015-09-18, -0.9, 0.2, -1.5, 0.1, 0.1, -1.6, -1.8, -1.5, -1.8, -1.7, -1.3; 2015-09-21, -0.5, 0.3, -1.8, 0.7, 0.7, -1.5, -2.3, -2.1, -2.4, -2.3, -2.1; 2015 -09-22, 1.3, 3.4, -0.4, -3.1,-3.1,1.7,0.8,0.9,0.9,0.7,1.0;2015-09-23,-0.8,0.7,-0.8,0.7,0.7,1.0, 0.1,0.3,0.2,-0.0,0.2; 2015-09-24,-0.1,1.5,-0.7,0.1,0.1,0.7,0.1,0.1,0.2,-0.1,0.1; 2015-09-25,1.5, 1.9,0.0,-0.9, -0.9,1.0, 1.0, 0.8, 1.0, 0.8, 0.9; 2015-09-28, 1.2, 1.2, -0.8, -0.2, -0.2, 0.5, 1.2, 1.0, 1.2, 0.9, 1.0; 2015-09-29, 1.3, 2.1, -0.3, -1.8, -1.8, 2.0, 3.0, 2.6, 2.8, 2.9, 2.8. \$codi评级为买入, 18.6\%华尔街分析师根据平均评级/目标得出的上涨。 \\ 
                          & Answer: \$codi的收盘价在2015-9-30会\textcolor{blue}{上涨}。 \\ 
    \hline
    CCIKM18 & Context: \$chk近期的走向趋势如下，date, open, high, low, close, adj-close, inc-5, inc-10, inc-15, inc-20, inc-25, inc-30; 2017-10-05，-0.7，0.7，-0.9，0.7，0.7，0.0，0.6，-1.3，-3.8，-5.6, -7.0; 2017-10-06, 1.4，1.7，-1.9, -2.3, -2.3, 2.0, 2.9, 1.4, -0.8，-2.8，-4.5; 2017-10-09，0.2，0.7, -1.4,-0.2,  -0.2, 1.4, 2.7, 1.9，0.1，-2.2，-3.9; 2017-10-10,7.9, 8.2, 0.0, -6.7, -6.7, 6.8, 8.9, 9.0, 7.5, 5.0，3.2; 2017-10-11, -0.8, 0.8, -4.8, 1.0, 1.0, 4.1, 6.6, 7.4, 6.4, 4.2, 2.4; 2017-10-12,1.3,2.1,-1.8,-2.5,-2.5，4.4，8.1，9.7，8.8，6.8，5.3; 2017-10-13, 0.5,1.3,-0.8, 0.8, 0.8，2.0，6.2，8.2，7.8，6.3，4.6。 \\ 
                          & Answer: \$chk的收盘价在2027-10-19会\textcolor{blue}{上涨}。 \\ 
    \hline
    CBigData22 & Context: \$msft近期的走向趋势如下，date, open, high, low, close, adj-close, inc-5, inc-10, inc-15, inc-20, inc-25, inc-30; 2020-10-09, -2.1, 0.0, -2.1, 2.5, 2.5, -2.5, -2.8, -3.6, -4.0, -4.0,-2.5 ; 2020-10-12, -1.2, 1.1, -2.1, 2.6, 2.6, -3.9, -4.7, -5.5, -6.1, -6.3, -5.1; 2020-10-13,-0.1,1.1,-1.1,0.7,0.7,-3.0, -4.6, -5.6, -6.4, -6.5, -5.8; 2020-10-14, 1.0, 1.5, -0.8, -0.9, -0.9, -1.2, -3.3,- 4.2, -5.2, -5.5, -5.0; 2020-10-15,-1.2, 0.3, -1.7, -0.5, -0.5,0.2,-2.4,-3.2,-4.3,-4.7,-4.7; 2020-10-16, 0.2, 1.2, -0.2, 0.0, 0.0, 0.6, -1.8, -2.8, -3.8, -4.5, -4.6; 2020-10-19, 2.9, 3.8, -0.2, -2.5, -2.5, 2.4, 0.9, -0.2, -1.1, -1.9, -2.2; 2020-10-20, 0.5, 1.3, -0.7, 0.2, 0.2, 1.5, 1.1, -0.2, -1.2, -2.0, -2.2; 2020-10-21,-0.8, 1.0, -0.8, 0.1, 0.1, 0.8, 1.2, -0.1, -0.9, -1.8, -2.2; 2020-10-22, -0.4, 0.5, -1.5, 0.0, 0.0, 0.4, 1.4, -0.1, -0.7, -1.7, -2.1。 \\ 
                          & Answer: \$msft的收盘价在2020-10-23会\textcolor{blue}{上涨}。 \\ 
    \hline
    ACL18 & Context: The recent trend of \$codi is as follows. date, open, high, low, close, adj-close, inc-5, inc-10, inc-15, inc-20, inc-25, inc-30; 2015-03-30, 0.0, 0.4, -0.9, 0.0, 0.0, -0.0, -0.8, -1.6, -2.1, -2.4, -2.6, 2015-03-31, 0.1, 0.3, -0.8, -0.3, -0.3, 0.2, -0.4, -1.0, -1.7, -2.0, -2.2; 2015-04-01, 1.1, 1.1, -0.7, -1.2, -1.2, 1.2, 0.9, 0.5, -0.4, -0.7, -0.9; 2015-04-02, -0.2, 0.4, -0.4, -0.2, -0.2, 1.0, 1.2, 0.7, -0.1, -0.4, -0.6; 2015-04-06, -0.6, 0.5, -0.9, 0.6, 0.6, 0.2, 0.6, 0.2, -0.6, -0.8, -1.1;2015-04-07, -0.4, 1.4, -0.4, 0.2, 0.2, -0.2, 0.4, 0.0, -0.5, -1.0, -1.2; 2015-04-08, 0.0, 0.2, -0.4, 0.1, 0.1, -0.4, 0.2, 0.0, -0.4,- 1.0, -1.2; 2015-04-09, -0.5, 0.2, -1.1, 0.6, 0.6, -0.7, -0.4, -0.5, -0.7, -1.4, -1.7; 2015-04-10, 0.1, 0.4, -0.5, 0.2, 0.2, -0.5, -0.6, -0.5,-0.8, -1.5, -1.7;2015-04-13, 1.4, 1.9, 0.0, -0.7, -0.7, 0.2, 0.1, 0.3, -0.0, -0.6, -0.9. \$codi - current report filing (8-k) 2015-04-13: william blair starts compass diversified \$codi at outperform. \\ 
                          & Answer: The closing price of \$codi will \textcolor{blue}{rise} at 2015-04-14. \\ 
    \hline
    CIKM18 & Context: The recent trend of \$aal is as follows. date, open, high, low, close, adj-close, inc-5, inc-10, inc-15, inc-20, inc-25, inc-30; 2017-01-17, 1.7, 2.1, -0.2, -1.9, -1.9, 2.5, 0.9, 1.3, 2.0, 2.3, 2.0; 2017-01-18, -0.5, 0.3, -2.1, 1.9, 1.9, 0.2, -0.7, -0.7, 0.1, 0.2, 0.2; 2017-01-19, 0.8, 1.8, -0.8, -0.8, -0.8, 0.5, 0.2, -0.1, 0.8, 1.0, 1.1; 2017-01-20, -1.0, 0.3, -1.6, 1.6, 1.6, -1.1, -0.9, -1.6, -0.9, -0.6, -0.3; 2017-01-23, 2.0, 2.4, -0.4, -2.2, -2.2,0.8,1.5,0.6,1.1,1.6,1.8 2017-01-24, -1.0, 0.5, -1.6, 1.3, 1.3, -0.1, 0.3, -0.5, -0.3, 0.2, 0.5; 2017-01-25,-0.0, 0.6, -0.6, 0.8, 0.8, -0.8, -0.6, -1.1, -1.1, -0.5, -0.4; 2017-01-26, -2.3, 0.2, -2.5, 3.5, 3.5, -3.2, -3.7, -4.1, -4.4, -3.8, -3.7 2017-01-27,6.5,6.5,-0.5,-5.3,-5.3, 1.8, 1.4, 1.5, 0.9, 1.4, 1.7; 2017-01-30, 1.6, 2.3, -2.3, -4.4, -4.4, 5.5, 5.5, 5.9, 5.3, 5.6, 6.1. \\ 
                          & Answer: The closing price of \$aal will \textcolor{blue}{fall} at 2020-9-28. \\ 
    \hline
    BigData22 & Context: The recent trend of \$intc is as follows. date, open, high, low, close, adj-close, inc-5, inc-10, inc-15, inc-20, inc-25, inc-30; 2020-09-14, -1.1, 0.6, -1.1, 0.3, 0.3, -0.4, 1.3, 1.0, 0.5, 0.1, -0.2; 2020-09-15, -0.4, 1.2, -0.5, 1.2, 1.2, -1.1, -0.1, -0.1, -0.6, -1.0, -1.3; 2020-09-16, 0.3, 1.3, -0.4, 0.7, 0.7, -1.5, -0.9, -0.7, -1.2, -1.5, -1.9; 2020-09-17, -1.9, 0.3, -2.0, -0.1, -0.1, -0.9, -1.2, -0.5, -0.9, -1.3, -1.7; 2020-09-18, 0.9, 1.2, -1.7, -0.9, -0.9, 0.2, -0.4, 0.4, 0.1,-0.4,-0.7; 2020-09-21, -0.7, 0.0, -1.8, -0.3, -0.3, 0.7, -0.1, 0.7, 0.5, 0.0, -0.3; 2020-09-22, -0.1, 0.5, -1.0, 0.5, 0.5, 0.2, -0.4, 0.1, 0.1, -0.4, -0.7; 2020-09-23, 2.1, 2.7, -0.3, -2.3, -2.3, 1.9, 1.7, 2.1, 2.3, 2.0, 1.6;2020-09-24, -1.3, 1.0, -1.5, 0.7, 0.7, 0.7, 1.1, 1.0, 1.6, 1.3, 0.9; 2020-09-25, -2.0, 0.7, -2.4, 1.6, 1.6, -0.8, -0.4, -0.6, 0.0, -0.2, -0.5. \\ 
                          & Answer: The closing price of \$intc will \textcolor{blue}{rise} at 2020-9-28. \\ 
    
    \hline
    \end{tabular}
  \label{tab:exa2}
\end{table*}

\begin{table*}[htb!]
    \centering
     \footnotesize
     \setlength\tabcolsep{3pt}
    \renewcommand{\arraystretch}{2.5}
    \caption{Examples with contexts and target answers in Event Detection(FinED), Relationship Extraction(FinRE), Text Summarization(FinTS), Entity Recognition(FinER), Credit Classification(FinCC) and Hawkish-dovish Classification(FinDC).}
    \begin{tabular}{m{2cm}<{\centering}|m{15cm}}
    \hline
    19CCKS & Context: 恒立实业(000622)遭证监会立案调查 涉嫌信披违规上海家化(600315)复星退出家化集团股权竞购。   \\ 
                          & Answer: 事件类型是\textcolor{blue}{信披违规}，事件主体为\textcolor{blue}{恒立实业}。 \\ 
    \hline
    20CCKS& Context: 2018年10月，中弘股份于发布公告称，截至2018年9月28日，公司逾期债务本息合计金额超过55亿元，全部为各类借款。   \\ 
                          & Answer: 事件类型是\textcolor{blue}{债务违约}，事件主体是\textcolor{blue}{中弘股份}。 \\ 
    \hline
    21CCKS & Context: 反观需求,由于生猪存栏量持续下降,而水产养殖启动缓慢,豆粕需求难以有效放大。供需失衡,导致国内外豆粕价格在豆类板块中呈现弱势。   \\ 
                          & Answer: 原因事件类型是\textcolor{blue}{供给减少}，原因实体是\textcolor{blue}{生猪}， 结果事件类型是\textcolor{blue}{市场价格下降}，结果实体是\textcolor{blue}{豆粕}。 \\ 
    \hline
    22CCKS & Context: 易投配资为什么无法出金；晨曦航空股东高文舍拟减持不超3\%股份；招行逾期第三个月的情况。   \\ 
                          & Answer: 事件类型是\textcolor{blue}{股东减持}，事件主体是\textcolor{blue}{晨曦航空}。 \\ 
    \hline
    RE & Context: 张裕与卡斯特集团合作的牵线人陈光因嫌分红太少而有意退出。实体对：张裕-卡斯特集团   \\ 
                          & Answer: 张裕和卡斯特集团的关系为\textcolor{blue}{合作}。 \\ 
    \hline
    NA & Context: 美今日凌晨，苹果发布iOS15.4首个测试版，更新后 Face ID将支持戴口罩解锁。 按照官方描述，在升级到戴口罩使用面容ID之后，开机就会提醒是否要使用此功能，用户选择使用后，会要求再录入一次面容ID，这次主要是记录使用者的眼睛周围特征，用于戴口罩时候识别。   \\ 
                          & Answer: 文本的摘要为\textcolor{blue}{苹果iOS15.4测试版支持戴口罩的面容解锁。}   \\ 
    \hline
    NER & Context: The proceeds of the equipment advances will be used solely to reimburse Borrower for the purchase of eligible equipment. \\ 
                          & Answer:  Entity name is \textcolor{blue}{Borrower}, and the corresponding entity type is \textcolor{blue}{PER}. \\ 
    \hline
    CNER & Context: 数、数年数月数日，银保监会核准王小林先生本行董事任职资格。   \\ 
                          & Answer:  实体名称是\textcolor{blue}{银保监会}, 类型是\textcolor{blue}{ORG}; 实体名称是\textcolor{blue}{王小林}, 类型是\textcolor{blue}{PER}。  \\ 
    \hline
    FINER-ORD& Context:   McPhail, 50, is angry and perplexed by politicians clamouring to sign a free trade deal with China that allows the importation of Chinese workers. \\ 
                          & Answer: The "token:label" labeling result for each row of data is \textcolor{blue}{[Bill:B-PER Shorten:I-PER pictured:O with:O the:O ALP:B-ORG 's:O candidate:O for:O Canning:B-LOC ,:O Matt:B-PER Keogh:I-PER .:O]} \\ 
    \hline
    ECTSUM & Context: And I'm confident that our disciplined approach to operating the business will result in our continued success throughout the balance of fiscal 2021. Coupled with the cost reduction actions we recently implemented, ..., Our Test business delivered a really solid Q1 by significantly beating our internal expectations and delivering an EBITDA margin of nearly 13\% versus 11\% last Q1.   \\ 
                          & Answer: The label whether a sentence is included in the abstract is \textcolor{blue}{[1, 0, 0, 0, 0, 0, 0, 0, 0, 0, 0]}. \\ 
    \hline
    EDTSUM & Context: ORLANDO, Fla., June 4, 2020 PRNewswire--Darden Restaurants, Inc., (NYSE: DRI) plans to release its fiscal 2020 fourth quarter financial results...For more information, please visit www.darden.com. Logo: https:mma.prnewswire.com media 340562 darden-restaurants-inc-logo.jpg SOURCE Darden Restaurants, Inc.: Financial Related Links http:www.Darden.com  \\ 
                          & Answer: The generated summary is \textcolor{blue}{Darden Restaurants to Host Fiscal 2020 Fourth Quarter Conference Call on June 25}. \\ 
    \hline
    German & Context: The client has attributes: X1: A14, X2: 12, X3: A32, X4: A40, X5: 1386, X6: A63, X7: A73, X8: 2, X9: A92, X10: A101, X11: 2, X12: A122, X13: 26, X14: A143, X15: A152, X16: 1, X17: A173, X18: 1, X19: A191, X20: A201.   \\ 
                          & Answer: The user's creditworthiness is \textcolor{blue}{bad}. \\ 
    \hline
    Australian & Context: The client has attributes: A1: 0.0, A2: 68.67, A3: 15.0, A4: 2.0, A5: 10.0, A6: 9.0, A7: 0.0, A8: 1.0, A9: 1.0, A10: 14.0, A11: 0.0, A12: 2.0, A13: 0.0, A14: 3377.0.  \\ 
                          & Answer: \textcolor{blue}{good} is the user's creditworthiness. \\ 
    \hline
    FOMC & Context: Nonetheless, given a recovery in U. S. domestic demand approximating their current forecasts, growth in imports likely would exceed that of exports by a wide margin over the forecast horizon.   \\ 
                          & Answer: The stance of this monetary policy text is \textcolor{blue}{hawkish}. \\ 
    \hline
    \end{tabular}
  \label{tab:exa3}
\end{table*}

\subsection*{D.  Human Evaluation for Expert-written Prompts }
\label{sec:appendixpro}

Table \ref{tab:eval1}and Table \ref{tab:eval2} illustrate examples of human evaluations for expert-written prompts across various financial-specific tasks.

\begin{table*}[htb!]
    \centering
    \footnotesize
     \setlength\tabcolsep{1pt}
    \renewcommand{\arraystretch}{2.5}
    \caption{Examples of human evaluation with metrics such as accuracy(AC), naturalness(NA), informativeness(IN) and and average(AVG) for domain-expert written prompts in various financial specific task. We retain the prompts with AVG above 2. }
    \begin{tabular}{m{2.8cm}<{\centering}|m{4cm}|m{4cm}|m{2cm}|m{2cm}|cccc}
    \toprule[1pt]
    \multirow{2}{=}{\centering\textbf{Specific Task}} & \multirow{2}{=}{\centering\textbf{Prompt}}   &  \multirow{2}{=}{\centering\textbf{Input}}   &  \multirow{2}{=}{\centering\textbf{Target}}  &  \multirow{2}{=}{\textbf{Response}}  &  \multicolumn{4}{c}{\textbf{Metric}}    \\ 
    &    &   &   &   &        AC & NA & IN & AVG \\
    \midrule
    \makecell{Sentiment \\ Analysis (FinSA)}  & 确定所提供的金融新闻文章中的句子的情绪，识别其中的情绪是积极，消极，还是中性的。只需回答'积极'、'中性'或'消极'。& 上涨空间异常巨大，中铁装配实在是未来的大牛股啊 &积极&积极&3&3 &3&3      \\
    \midrule
    \makecell{Sentiment \\ Matching (FinSM)}   & 你需要判断给出的两个金融文本表达的意思是否相似，请回答是或否。&  花呗咋扫不上 / 花呗扫一扫不能付钱 &是&否&0&1&1&0.7   \\
    \midrule
    \makecell{News Classification \\ (FinNC)}     & 请根据金融报道的内容，判断其涉及范围属于'中国'、'国际'还是'外国'？ &  伯恩斯坦：首予康菲石油(COP.N)跑赢大盘评级，目标价为98美元。 &外国&国际&1&3&1&1.7    \\
    \midrule
    \makecell{Negative Judgment \\ (FinNJ)}   & 根据所给的金融新闻及实体，你需要判断所给实体是否含有负面的消息，你只需简单回答'有'或者'无'。   &  自从号称750亿交易量的唐小僧倒下...几乎每天都有几家网贷公司宣布清盘、提现困难、经侦介入等。实体：唐小僧;多米 &有&有&3&3&3&3      \\
    \midrule
    \makecell{Answer Selection \\ (FinAS)}       &  根据所提出的金融问题，请从以下四个选项中选择最合适的一个。你的输出应该是：'A'、'B'、'C' 或'D'。 &  如果一个竞争性市场位于长期均衡状态中，那么所有的厂商()。 A,采用完全相同的生产工艺; B,具有同一最低平均的成本; C,都能获得经济利润; D,以上全对 &D&D&3&3&2&2.7    \\
    \midrule
    \makecell{Stock Prediction \\ (FinSP)}     & 请根据新闻对股票数据的影响判断该公司的股票运动走势是跑赢大盘、中性还是表现不佳。 &  高新兴公告，实控人拟减持不超过6\%。日期 开盘价 闭盘价，2020-09-23, 5.56, 5.52;2020-09-24, 
     5.47, 5.22; ... ; 2020-09-29, 5.02, 5.03 &表现不佳&无法预测&0&1&2&1       \\
    \midrule
    \makecell{Headline Classification \\ (FinHC)}&  下面所给内容是否提到了黄金商品的上涨趋势？请回答是或否。   &  金价在印度的主要城市上涨 &是&是&3&3&2&2.7  	     \\
    \midrule
    \makecell{Question Answering \\ (FinQA)}    &  你需要分析金融文本，根据内容回答相关问题。   & 【亿邦动力讯】10月21日消息，洛丁森完成B轮融资，...，下文中发生的事件类型有哪些? &企业融资&无法确定&0&0&0&0     \\
    \midrule
    \makecell{Event Detection\\(FinED)}       & 阅读以下金融领域的公告，判断所有的事件类型及其对应主体。请以:'事件类型, 事件主体'的格式回复。其中事件类型应该在['信批违规'，...，'涉嫌欺诈']中。  &  劲嘉股份(002191)因虚假陈述遭证监会处罚浩云科技(300448)股东披露减持计划 &信批违规, 劲嘉股份&信批违规, 劲嘉股份&3&3&3&3    \\
    \midrule
    \makecell{Entity Recognition \\ (FinER)}     &  在下面所给金融语句中，识别指明个人('PER')、组织('ORG')或地点('LOC')的特定命名实体。  &  报告期内，本公司除中国银保监会批准的经营范围内的金融担保业务外，没有其他需要披露的重大担保事项。 &中国银保监会, ORG&中国银保监会, ORG;中国, LOC&2&2&3&2.3   \\
    \midrule
    \makecell{Relationship Extraction\\ (FinRE)} &  请仔细分析所给金融报道和实体对，然后在['合并', '竞争',...]中选择能准确描述该实体对关系的选项。 &  张裕与卡斯特集团合作的牵线人陈光因嫌分红太少而有意退出。实体对：卡斯特集团-张裕 &合作&合作关系受损&1&2&3&2                      \\
    \midrule
    \makecell{Text Summarization \\ (FinTS)}     &  你的任务是对所给的金融文本进行一句话摘要,要求尽量简短。 &  前段时间，佳能宣布关闭经营了32年的珠海工厂，引发网络关注。...而优厚的补偿，也是佳能自愿选择的“体面离开”。（新京报）&佳能撤离珠海，不宜说成“恶意赔偿”&佳能关闭珠海工厂，提供优厚补偿，引发争议。&1&2&3&2    \\  
    \bottomrule[1pt]
    \end{tabular}
    \label{tab:eval1}
\end{table*}

\begin{table*}[htb!]
    \centering
   \footnotesize
     \setlength\tabcolsep{1pt}
    \renewcommand{\arraystretch}{1}
    \caption{Examples of human evaluation with metrics such as accuracy(AC), naturalness(NA), informativeness(IN) and and average(AVG) for domain-expert written prompts in various financial specific task. We retain the prompts with the AVG metric above 2.}
    \begin{tabular}{m{2.8cm}<{\centering}|m{4cm}|m{4cm}|m{2cm}|m{2cm}|cccc}
    \toprule[1pt]
    \multirow{2}{=}{\centering\textbf{Specific Task}} & \multirow{2}{=}{\centering\textbf{Prompt}}   &  \multirow{2}{=}{\centering\textbf{Input}}   &  \multirow{2}{=}{\centering\textbf{Target}}  &  \multirow{2}{=}{\textbf{Response}}  &  \multicolumn{4}{c}{\textbf{Metric}}    \\ 
    &    &   &   &   &        AC & NA & IN & AVG \\
    \midrule
    \makecell{Sentiment  Analysis\\ (FinSA)}  & Determine the mood of the sentences in the financial news article provided, identifying whether the mood in them is positive, negative, or neutral. Simply answer 'positive', 'neutral', or 'negative'. & Up space is exceptionally huge, China Railway assembly, is really the future of the big bull stock ah &positive&positive&3&3 &3&3      \\
    \midrule
    \makecell{Sentiment  Matching\\ (FinSM)}   & You need to decide whether the two financial texts given express similar meanings, answer yes or no.&  Why can't I pay/ I can't pay by scanning. &yes&no&0&1&1&0.7   \\ 
    \midrule 
    \makecell{News Classification \\ (FinNC)}     & Based on the content of the financial report, please determine whether its scope is 'Chinese', 'international' or 'foreign'? &Bernstein: first Outperform rating on ConocoPhillips (COP.N), \$98 price target. &foreign&international&1&3&1&1.7    \\
    \midrule
    \makecell{Negative Judgment \\ (FinNJ) }  & Based on the given financial news and entity, you need to determine whether the given entity contains negative news or not, you simply need to answer 'yes' or 'no'.   &Since the fall of Tang Xiaosheng, which claimed a transaction volume of 75 billion ... Almost every day, there are several online lending companies announced liquidation, cash withdrawal difficulties, the investigation intervention, etc. Entity: Tang Xiaosheng; Domi &yes&yes&3&3&3&3      \\
    \midrule
    \makecell{Answer Selection \\ (FinAS) }      &  Based on the financial question posed, choose the most appropriate of the following four options. Your output should be: 'A', 'B', 'C' or 'D'. &If a competitive market is in long-run equilibrium, then all firms () A, use exactly the same production process; B, have the same minimum average cost; C, are economically profitable; D, all of the above. &D&D&3&3&2&2.7    \\
    \midrule
    \makecell{Stock Prediction\\ (FinSP)}     & Determine whether the company's stock movement trend is outperforming, neutral, or underperforming based on the impact of the news on the stock data. &  Gaoxin announced that the real controller intends to reduce its holdings by no more than 6\%. Date  Opening price  Closing price : 2020-09-23, 5.56, 5.52 ; 2020-09-24, 5.47, 5.22; ... ; 2020-09-29, 5.02, 5.03 &underperformance&unpredictable&0&1&2&1       \\
    \midrule
    \makecell{Headline Classification \\ (FinHC)}&  Does the content given below refer to an upward trend in gold commodities? Please answer yes or no.   &  Gold prices rise in India's major cities &yes&yes&3&3&2&2.7  	     \\
    \midrule
    \makecell{Question Answering \\ (FinQA)}    &  You will need to analyze financial texts and answer relevant questions based on the content.   & [YPC] October 21, 2011 - Lodgingsen closed its Series B financing... What are the types of events that occur in the following article? &corporate financing&indeterminate&0&0&0&0     \\
    \midrule
    \makecell{Event Detection\\ (FinED)}       & Read the following announcements in the financial sector and determine all event types and their corresponding subjects. Please respond in the format :'Event Type, Event Subject'. Where the event type should be in ['credit approval violation', ...' Suspected Fraud'].  &  Jinjia shares (002191) due to misrepresentation by the Securities and Futures Commission penalized haoyun science and technology (300448) shareholders disclosure of shareholding reduction plan &Letter approval violation, Jinjia shares&Letter approval violation, Jinjia shares&3&3&3&3    \\
    \midrule
    \makecell{Entity Recognition \\(FinER)}     &  In the financial statement given below, identify the specific named entity that specifies a person ('PER'), organization ('ORG'), or location ('LOC').。  &  During the reporting period, the Company did not have any other material guarantee matters requiring disclosure, except for the financial guarantee business within the scope of operation approved by the China Banking and Insurance Regulatory Commission. &China Banking and Insurance Regulatory Commission, ORG&China Banking Regulatory Commission, ORG; China, LOC&2&2&3&2.3   \\
    \midrule
    \makecell{Relationship Extraction\\ (FinRE)} &  Carefully analyze the given financial reports and entity pairs, and then among ['Merger', 'Competition', ...] Choose the option that accurately describes the relationship of the entity pair. &  Chen Guang, the matchmaker of Zhangyu's partnership with Custer Group, intends to quit because he thinks the dividend is too small. Entity Pair: Custer Group - Zhang Yu & collaborate&Impaired partnership&1&2&3&2                      \\
    \midrule
    \makecell{Text Summarization\\ (FinTS)}     & Your task is to make a one-sentence summary of the given financial text, keeping it as short as possible. &  Some time ago, Canon announced the closure of its Zhuhai plant, which had been in operation for 32 years, triggering concern on the Internet.... The generous compensation is also Canon's voluntary choice to "leave with dignity". (New Beijing News)&Canon's withdrawal from Zhuhai should not be labeled as "malicious compensation".&Canon closes Zhuhai plant, offers generous compensation, sparks controversy.&1&2&3&2    \\  
    \bottomrule[1pt]
    \end{tabular}
\label{tab:eval2}
\end{table*}

\subsection*{E. The Detailed Evaluation Results of ICE-FLARE}
\label{sec:appendixres}

Table \ref{tab:res} presents the detailed evaluation results of 6B-7B general LLMs, baseline FinLLMs, and our proposed ICE-INTERN variants.

\begin{table*}[htb!]
\centering
    \footnotesize
    \setlength\tabcolsep{1pt} 
   \renewcommand{\arraystretch}{0.6}
   \caption{The detailed evaluation results of different 6B-7B LLMs on ICE-FLARE.  “-” represents the result that is currently unable to yield due to model size or availability. Results with "*" of ChatGPT(3.5-turbo-1106) and GPT-4 on English datasets are cited from ~\cite{xie2023pixiu}. Results in bold indicate the best results across all models. Results with "\_\_" indicate the best results among models excluding ChatGPT.}
    \begin{tabular}{ccccccccccccccccccccccc}
    \toprule[1pt]
    \rotatebox[origin=c]{90}{\textbf{Language}} & \rotatebox[origin=c]{90}{\textbf{Data Type}} & \rotatebox[origin=c]{90}{\textbf{NLP Task}} & \rotatebox[origin=c]{90}{\textbf{Specific Task}} &  \rotatebox[origin=c]{90}{\textbf{Metric}} & \rotatebox[origin=c]{90}{\textbf{Dataset}} & \rotatebox[origin=c]{90}{\textbf{Bloomz-7B1}} & \rotatebox[origin=c]{90}{\textbf{Baichuan-7B}} & \rotatebox[origin=c]{90}{\textbf{Baichuan2-7B}} & \rotatebox[origin=c]{90}{\textbf{LLaMa2-7B}} & \rotatebox[origin=c]{90}{\textbf{ChatGLM2-6B}} & \rotatebox[origin=c]{90}{\textbf{ChatGLM3-6B}} & \rotatebox[origin=c]{90}{\textbf{Qwen-7B}} & \rotatebox[origin=c]{90}{\textbf{ChatGPT}} & \rotatebox[origin=c]{90}{\textbf{GPT-4}} & \rotatebox[origin=c]{90}{\textbf{InternLM-7B}} &\rotatebox[origin=c]{90}{\textbf{CFGPT-7B}}& \rotatebox[origin=c]{90}{\textbf{DISCFin-13B}}&  \rotatebox[origin=c]{90}{\textbf{FinMA}}&\rotatebox[origin=c]{90}{\textbf{ICEdlc-7B}} & \rotatebox[origin=c]{90}{\textbf{ICEdle-7B}} & \rotatebox[origin=c]{90}{\textbf{ ICEdtt-7B}} & \rotatebox[origin=c]{90}{\textbf{ICEfull-7B}} \\ 
    
    \midrule
    \multirow{30}{*}{\rotatebox[origin=c]{90}{ZH}} & \multirow{22}{*}{\rotatebox[origin=c]{90}{DLC}} & \multirow{22}{*}{ZH-CLS} & \multirow{3}{*}{FinSA} & \multirow{2}{*}{Accuracy} 
             & FE   & 0.587 & 0.442 & 0.002 & 0.423 & 0.000 & 0.572 & 0.366 & 0.329 & 0.349 & 0.553 & 0.306 & 0.624 & 0.534 & 0.760 & \textbf{0.778} & 0.743  &0.760    \\
              &  &            &                        &                          & StockB & 0.196 & 0.058 & 0.000 & 0.047 & 0.000 & 0.389 & 0.431 & 0.310 & 0.441 & 0.297 & 0.070 & 0.282 & 0.202 & 0.422 & 0.593 & \textbf{0.636} &0.593    \\
              &  &           &                        & F1                         & StockB & 0.177 & 0.098 & 0.000 & 0.078 & 0.000 & 0.443 & 0.482 & 0.313 & 0.462 & 0.279 & 0.098 & 0.344 & 0.201 & 0.395 & 0.571 & 0.575 &\textbf{0.578}    \\
    
    \cmidrule(lr){4-23}
              &  &           & \multirow{2}{*}{FinSM} & \multirow{2}{*}{F1} & BQC & 0.008 & 0.154 & 0.002 & 0.292 & 0.002 & 0.444 & 0.334 & 0.682 &  0.704 & 0.278& 0.218 & 0.255 & 0.230  &  0.854  & \textbf{0.859} &0.846 &0.854     \\
               &  &           &                        &                           & AFQMC  & 0.004 & 0.296 & 0.014 & 0.266 & 0.012 & 0.470 & 0.560 & 0.323 &  0.391 & 0.292& 0.460 & 0.373 & 0.095  &  0.743  & 0.609 &0.594 &\textbf{0.743}     \\
    \cmidrule(lr){4-23}
              &  &           & \multirow{2}{*}{FinNC} & \multirow{2}{*}{F1} & NL & 0.245 & 0.219 & 0.002 & 0.280 & 0.000 & 0.453 & 0.348 & 0.645 &  0.816 & 0.375& 0.164 & 0.267 & 0.261  &  0.950  & 0.922 &\textbf{0.954} &0.950    \\
              &  &           &                        &                           & NL2  & 0.000 & 0.049 & 0.000 & 0.066 & 0.000 & 0.049 & 0.074 & 0.034 &  0.273 & 0.039& 0.004 & 0.002 & 0.060  &  0.786  & 0.774 &0.768 &\textbf{0.786}     \\
    \cmidrule(lr){4-23}
              &  &           & FinNJ                  & F1 &                NSP & 0.041 & 0.217 & 0.000 & 0.353 & 0.004 & 0.558 & 0.636 & \textbf{0.935} &  0.916 & 0.351& 0.435 & 0.208 & 0.442  &  0.692  & 0.692 &\underline{0.695} &0.692    \\
    \cmidrule(lr){4-23}
               &  &          & \multirow{2}{*}{FinAS}  & Accuracy &                FinevalF & 0.302 & 0.329 & 0.014 & 0.266 & 0.185 & 0.423 & 0.243 & 0.491 &  \textbf{0.662} & 0.320 & 0.284 & 0.369 & 0.261 &  0.441  & 0.410 &0.387 &\underline{0.441}    \\
               &  &          &                         & F1 &               FinevalF & 0.295 & 0.319 & 0.024 & 0.171 & 0.190 & 0.418 & 0.096 & 0.481 &  \textbf{0.686 }& 0.247& 0.264 & 0.362 & 0.220  &  0.442  & 0.387 &0.376 &\underline{0.442}    \\
    \cmidrule(lr){4-23}
              &   &          & FinRE                  & F1 &                RE & 0.012 & 0.039 & 0.000 & 0.011 & 0.000 & 0.076 & 0.026 & 0.211 &  \textbf{0.340} & 0.015& 0.014 & \underline{0.108} & 0.009  &  0.092  & 0.088 &0.103 &0.103    \\
    \cmidrule(lr){3-23}
             &     & ZH-PRE     & FinSP                  & F1 &                StockA & 0.131 & 0.000 & 0.000 & 0.000 & 0.000 & 0.220 & 0.154 & 0.422 &  0.383 & 0.003 & 0.050 & 0.315 & 0.000 &  0.340  & \textbf{0.637} &0.630 &0.617    \\
    \cmidrule(lr){2-22}
    & \multirow{13}{*}{\rotatebox[origin=c]{90}{DLE}} & \multirow{10}{*}{ZH-EXT} & FinQA  & Accuracy &  QA & 0.098 & 0.001 & 0.000 & 0.001 & 0.000 & 0.099 & 0.000 & 0.256 &  0.064 & 0.001 & 0.092 & 0.201 & 0.123 &  0.167  & \textbf{0.864} &0.860 &0.855    \\
    \cmidrule(lr){4-23}
              &  &          & FinER                  & Entity F1 &                CNER & 0.000 & 0.000 & 0.000 & 0.000 & 0.000 & 0.000 & 0.000 & 0.101 &  0.103 & 0.000 & 0.000 & 0.000 & 0.000 &  0.000  & \textbf{0.483} &0.482 &0.468    \\
    \cmidrule(lr){4-23}
             &  &          & \multirow{8}{*}{FinED} & \multirow{4}{*}{Precision} & 19CCKS & 0.080 & 0.013 & 0.003 & 0.001 & 0.012 & 0.001 & 0.001 & 0.010 &  0.011 & 0.003 & 0.003 & 0.004 & 0.004 &  0.005  & \textbf{0.831} &0.825 &0.824    \\
             &  &          &                        &                           & 20CCKS  & 0.052 & 0.006 & 0.002 & 0.001 & 0.009 & - & 0.001 & 0.001 &  0.008 & 0.002 & 0.002 & 0.002 & 0.003 &  0.006  & 0.740 &0.741 &\textbf{0.743}     \\
             &  &          &                        &                           & 21CCKS  & 0.000 & 0.000 & 0.000 & 0.000 & 0.000 & 0.000 & 0.000 & 0.001 &  \textbf{0.004} & 0.000 & 0.000 & 0.000 & 0.000 &  0.000  & 0.001 &0.001 &\underline{0.002 }    \\
             &  &          &                        &                           & 22CCKS  & 0.000 & 0.001 & 0.001 & 0.000 & 0.002 & 0.001 & 0.000 & 0.006 &  - & 0.001 & 0.003 & 0.000 & 0.021 &  0.007  & 0.647 &0.642 &\textbf{0.660}     \\
             &  &          &                         & \multirow{4}{*}{F1}&    19CCKS & 0.081 & 0.013 & 0.003 & 0.001 & 0.012 & 0.001 & 0.001 & 0.010 &  0.012 & 0.003& 0.006 & 0.005 & 0.007  &  0.005  & \textbf{0.831} &0.825 &0.824    \\
             &  &          &                        &                           & 20CCKS  & 0.068 & 0.011 & 0.002 & 0.003 & 0.015 & - & 0.002 & 0.001 &  0.010 & 0.003& 0.004 & 0.004 & 0.005  &  0.004  & \textbf{0.705} &0.703 &0.702     \\
             &  &         &                        &                           & 21CCKS  & 0.000 & 0.000 & 0.000 & 0.000 & 0.000 & 0.000 & 0.000 & 0.001 &  \textbf{0.003} & 0.000 & 0.000 & 0.000 & 0.000 &  0.000  & 0.001 &0.001 &\underline{0.001}     \\
             &  &          &                        &                           & 22CCKS  & 0.001 & 0.001 & 0.001 & 0.001 & 0.004 & 0.001 & 0.001 & 0.008 &  - & 0.001 & 0.004 & 0.000 & 0.002 &  0.007  & 0.588 &0.581 &\textbf{0.595}     \\ 
    \cmidrule(lr){3-23}                     
             &  &       \multirow{3}{*}{ZH-GEN}     & \multirow{3}{*}{FinTS}                 & Rouge-1                     & NA & 0.078 & 0.028 & 0.006 & 0.104 & 0.001 & 0.256 & 0.142 & 0.093 &  - & 0.145& 0.006 & 0.014 & 0.010  &  0.050  & 0.284 &\textbf{0.289} &0.287    \\
             &  &          &                       & BertScore                 & NA & 0.676 & 0.527 & 0.482 & 0.649 & 0.496 & 0.748 & 0.603 & 0.710 &  - & 0.651& 0.603 & 0.607 & 0.603  &  0.707  & 0.824 &0.823 &\textbf{0.824}    \\
             &  &          &                       & BartScore                & NA & -5.704 & -6.845 & -7.274 & -4.539 & -7.293 & -4.097 & -4.301 & -5.882 &  - & -4.153 & -6.661 & -6.194 & -6.640 &  -6.044  & -4.180 &-4.183 &\textbf{-4.027}    \\
    \cmidrule(lr){2-23}
    &  \multirow{12}{*}{\rotatebox[origin=c]{90}{DTT}} & \multirow{12}{*}{ZH-TRA} & \multirow{3}{*}{FinSA} & Accuracy & CFPB&  0.005 & 0.203 & 0.000 & 0.098 & 0.000 & 0.424 & 0.239 & 0.699 &  0.196 & 0.273& 0.256 & 0.085 & 0.038  &  0.545  & 0.498 &0.846 &\textbf{0.849} \\
              & &           &          &\multirow{2}{*}{F1} & CFPB   & 0.010 & 0.129 & 0.000 & 0.092 & 0.000 & 0.348 & 0.340 & 0.642 &  0.306 & 0.209& 0.346 & 0.091 & 0.067  &  0.529  & 0.485 &0.845 &\textbf{0.846}     \\
             &  &           &                        &                          & CFiQA-SA & 0.652 & 0.401 & 0.000 & 0.350 & 0.000 & 0.617 & 0.773 & 0.400 &  0.120 & 0.668& 0.420 & 0.784 & 0.616  &  0.752  & 0.709 &0.796 &\textbf{0.868}    \\
    \cmidrule(lr){4-23}
              & &          & \multirow{6}{*}{FinSP}  & \multirow{3}{*}{Accuracy} & CACL18 & 0.550 & 0.585 & \textbf{0.591} & 0.577 & 0.589 & 0.485 & 0.556 & 0.544 &  0.568 & 0.566& 0.581 & 0.564 & 0.585  &  0.438  & 0.442 &0.581 &0.417    \\
               & &          &                        &                           & CBigData18  & 0.667 & 0.686 & 0.686 & 0.686 & 0.686 & 0.440 & 0.673 & 0.629 &  0.642 & \textbf{0.686}& 0.535 & 0.585 & 0.660  &  0.365  & 0.403 &0.660 &0.321     \\
               & &          &                        &                           & CCIKM18   & 0.384 & 0.372 & 0.372 & 0.372 & 0.372 & 0.605 & 0.384 & 0.326 &  0.419 & 0.372& 0.407 & 0.477 & 0.361  &  0.605  & \textbf{0.628} &0.395 &0.605    \\   
              & &          &                         & \multirow{3}{*}{MCC}  & CACL18 & -0.034 & 0.008 & 0.000 & 0.021 & -0.002 & 0.004 & 0.012 & 0.002 &  0.010 & -0.017& \textbf{0.099} & 0.076 & 0.022  &  0.044  & 0.042 &0.039 &-0.034    \\
              & &          &                        &                           & CBigData18  & -0.013 & 0.000 & 0.000 & 0.000 & 0.000 & -0.042 & -0.023 & -0.024 &  0.066 & 0.000& 0.067 & 0.037 & -0.109  &  0.028  & -0.056 &\textbf{0.114} &-0.033     \\
              &  &          &                        &                           & CCIKM18   & 0.084 & 0.000 & 0.000 & 0.000 & 0.000 & 0.133 & 0.084 & -0.250 &  0.032 & 0.000 & -0.066 & 0.040 & -0.141 &  0.037  & \textbf{0.204} &-0.021 &-0.119     \\
    \cmidrule(lr){4-23}
               & &          & FinHC                 &  AvgF1                  & CHeadlines & 0.503 & 0.500 & 0.500 & 0.500 & 0.500 & 0.695 & 0.500 & 0.769 &  0.818 & 0.504& 0.607 & 0.716 & 0.626  &  0.690  & 0.526 &0.952 &\textbf{0.957}    \\
    \cmidrule(lr){4-23}
               & &          & \multirow{2}{*}{FinQA}  & EM  & CEnQA & 0.000 & 0.000 & 0.000 & 0.000 & 0.000 & 0.000 & 0.000 & 0.000 &  0.000 & 0.000 & 0.000 & 0.000 & 0.000 &  0.000  & 0.000 &0.000 &\textbf{0.008}    \\
               & &          &                         & Accuracy &                        ConFinQA & 0.068 & 0.004 & 0.000 & 0.008 & 0.000 & 0.021 & 0.000 & 0.000 &  0.021 & 0.000 & 0.008 & 0.030 & \textbf{0.276}  &  0.013  & 0.046 &0.169 &0.228    \\
    \midrule
    \midrule
    \multirow{26}{*}{\rotatebox[origin=c]{90}{EN}} & \multirow{17}{*}{\rotatebox[origin=c]{90}{DTE}} & \multirow{8}{*}{EN-CLS} &\multirow{4}{*}{FinSA} & Accuracy & FPB   & 0.388 & 0.231 & 0.000 & 0.286 & 0.001 & 0.741 & 0.696 & 0.780* &  0.760* & 0.319 & 0.278 & 0.808 & 0.435 &  0.353  & 0.384 &0.398 &\textbf{0.867 }    \\
               & &         &                         & \multirow{2}{*}{F1}  & FPB   & 0.231 & 0.171 & 0.000 & 0.127 & 0.002 & 0.744 & 0.702 & 0.780* &  0.780* & 0.353& 0.292 & 0.809 & 0.436  &  0.349  & 0.331 &0.415 &\textbf{0.866}     \\
               & &         &                        &                          & FiQA-SA & 0.742 & 0.327 & 0.000 & 0.370 & 0.000 & 0.551 & 0.754 & 0.600* &  0.800* & 0.349& 0.096 & 0.532 & 0.477  &  0.477  & 0.503 &0.480 &\textbf{0.848}    \\
    \cmidrule(lr){4-23}
               & &           & FinHC                 &  AvgF1                  & Headlines & 0.600 & 0.600 & 0.600 & 0.600 & 0.600 & 0.655 & 0.600 & 0.770* &  0.860* & 0.600& 0.551 & 0.600 & 0.600  &  0.682  & 0.668 &0.780 &\textbf{0.965}    \\
    \cmidrule(lr){4-23}
               & &          & \multirow{4}{*}{FinCC}  & \multirow{2}{*}{F1} & German & \textbf{0.649} & 0.525 & 0.525 & 0.525 & 0.525 & 0.525 & 0.476 & 0.200* &  0.550* & 0.525& 0.525 & 0.548 & 0.192  &  0.525  & 0.525 &0.525 &0.318    \\
                & &         &                         &                     &Australian & 0.412 & 0.260 & 0.260 & 0.260 & 0.358 & 0.273 & 0.502 & 0.410* &  \textbf{0.740}* & 0.260& 0.288 & 0.260 & \underline{0.599}  &  0.260  & 0.260 &0.260 &0.421    \\
                & &         &                         & \multirow{2}{*}{MCC} & German & \textbf{0.231} & 0.000 & 0.000 & 0.000 & 0.000 & 0.000 & -0.166 & -0.100* &  -0.020* & 0.000& 0.000 & 0.012 & 0.002  &  0.000  & 0.000 &0.000 &-0.306    \\
                & &         &                         &                      &Australian & 0.000 & 0.000 & 0.000 & 0.000 & 0.153 & -0.017 & 0.065 & 0.000* &  \textbf{0.470}* & 0.000& -0.097 & 0.000 & \underline{0.232}  &  0.000  & 0.000 &0.000 &-0.030    \\
    \cmidrule(lr){3-23}
              & &  \multirow{8}{*}{EN-PRE}          & \multirow{6}{*}{FinSP}  & \multirow{3}{*}{Accuracy} & ACL18 & 0.503 & 0.500 & 0.487 & 0.508 & 0.488 & 0.511 & 0.490 & 0.500* &  0.520* & 0.510& 0.483 & 0.492 & 0.493  &  0.515  & 0.519 &\textbf{0.525} &0.502    \\
              & &           &                        &                           & BigData18  & 0.550 & 0.525 & 0.448 & 0.484 & 0.448 & 0.4769 & 0.508 & 0.530* &  0.540* & 0.469& 0.452 & 0.472 & 0.496  &  \textbf{0.555}  & 0.520 &0.483 &0.450     \\
              & &           &                        &                           & CIKM18   & 0.460 & 0.478 & 0.419 & 0.496 & 0.418 & 0.499 & 0.470 & 0.550* &  \textbf{0.570}* & 0.469& 0.421 & 0.430 & 0.461  &  \underline{0.563}  & 0.556 &0.534 &0.493     \\
              & &           &                         & \multirow{3}{*}{MCC}  & ACL18 & -0.034 & -0.010 & 0.000 & 0.010 & 0.023 & 0.035 & -0.038 & 0.005* &  0.020* & 0.019& -0.026 & 0.001 & -0.014  &  0.017  & 0.029 &\textbf{0.045} &0.023    \\
              & &           &                        &                           & BigData18  & 0.002 & -0.013 & 0.000 & -0.012 & 0.000 & 0.008 & -0.017 & -0.025* &  0.030* & -0.014& 0.030 & \textbf{0.063} & -0.002  &  0.042  & 0.030 &0.017 &-0.012     \\
               & &          &                        &                           & CIKM18   & -0.053 & 0.009 & 0.000 & -0.028 & -0.026 & \textbf{0.043} & -0.056 & 0.010* &  0.020* & -0.034& -0.016 & -0.044 & -0.038  &  -0.001  & -0.007 &0.005 &0.042     \\
     \cmidrule(lr){3-23}                    
         & &  EN-EXT  & FinER  & Entity F1 & NER & 0.000 & 0.000 & 0.000 & 0.012 & 0.000 & 0.248 & 0.012 & 0.770* &  \textbf{0.830}* & 0.000& 0.000 & 0.039 & \underline{0.392}  &  0.000  & 0.006 &0.000 &0.362    \\

    \cmidrule(lr){3-23}
           & &  \multirow{2}{*}{EN-REA}   & \multirow{2}{*}{FinQA}  &  EM & EnQA & 0.010 & 0.000 & 0.000 & 0.000 & 0.000 & 0.000 & 0.000 & 0.580* &  \textbf{0.630}* & 0.000& 0.000 & \underline{0.004} & 0.001  &  0.000  & 0.000 &0.001 &0.001    \\
           & &             &                         & Accuracy  &                        ConFinQA & 0.000 & 0.000 & 0.000 & 0.000 & 0.000 & 0.000 & 0.000 & 0.600* &  \textbf{0.760}* & 0.000 & \underline{0.001} & 0.000 & 0.000 &  0.000  & 0.000 &0.000 &0.000    \\
    \cmidrule(lr){2-23}
           & \multirow{9}{*}{\rotatebox[origin=c]{90}{DOT}} &  \multirow{9}{*}{EN-DOT}         & FinER & Entity F1 &  FINER-ORD & 0.000 & 0.000 & 0.000 & 0.000 & 0.000 & \underline{0.021} & 0.000 & 0.280* &  \textbf{0.770}* & 0.000& 0.000 & 0.008 & 0.000  &  0.000  & 0.000 &0.000 &0.000    \\
    \cmidrule(lr){4-23}
            &&             & \multirow{6}{*}{FinST}  & \multirow{2}{*}{Rouge-1} & ECTSUM & 0.000 & 0.000 & 0.000 & 0.000 & 0.000 & 0.000 & 0.000 & 0.000 &  0.000 & 0.000 & 0.000 & 0.000 & 0.000 &  0.000  & 0.000 &0.000 &\textbf{0.000}    \\
             &&            &                         &  &                        EDTSUM & 0.013 & 0.015 & 0.007 & 0.027 & 0.009 & \underline{0.175} & 0.036 & 0.170* &  \textbf{0.200}* & 0.032& 0.008 & 0.051 & 0.033  &  0.009  & 0.018 &0.017 &0.015    \\
             &&            &                          & \multirow{2}{*}{BertScore} & ECTSUM & 0.000 & 0.000 & 0.000 & 0.000 & 0.000 & 0.000 & 0.000 & 0.000 &  0.000 & 0.000  &  0.000  & 0.000 &0.000 &0.000  & 0.000 &0.000 &\textbf{0.000}    \\
             & &           &                         &  &                        EDTSUM & 0.522 & 0.447 & 0.445 & 0.483 & 0.445 & \underline{0.599} & 0.536 & 0.660* &  \textbf{0.670}* & 0.478& 0.508 & 0.552 & 0.467  &  0.476  & 0.536 &0.505 &0.500    \\
            &&             &                         & \multirow{2}{*}{BartScore} & ECTSUM & -5.178 & -5.178 & -5.178 & -5.178 & -5.178 & -5.178 & -5.178 & -5.178* &  -5.178* & -5.178 & -5.178 & -5.178 & -5.178 &  -5.178  & -5.178 &-5.178 &\textbf{-5.178}    \\
            &&            &                         &  &                        EDTSUM & -6.996 & -6.923 & -7.036 & -6.902 & -6.957 & \underline{-4.048} & -6.845 & -3.640* &  \textbf{-3.620}* & -6.973& -7.083 & -6.951 & -6.964  &  -7.107  & -6.945 &-6.925 &-6.927    \\
    \cmidrule(lr){4-23}
          && & \multirow{2}{*}{FinDC}  &                   Accuracy & FOMC & 0.288 & 0.256 & 0.000 & 0.319 & 0.000 & 0.464 & 0.454 & 0.600* &  \textbf{0.690}* & 0.252& 0.208 & \underline{0.557} & 0.476  &  0.385  & 0.373 &0.488 &0.429    \\
           &&              &                         &   F1             &FOMC & 0.207 & 0.172 & 0.000 & 0.323 & 0.000 & 0.470 & 0.441 & 0.640* &  \textbf{0.710}* & 0.131& 0.155 & \underline{0.541} & 0.480  &  0.375  & 0.372 &0.455 &0.424    \\
    \midrule
    \multicolumn{3}{c}{\multirow{2}{*}{Best Result Count}}  & \multicolumn{3}{c}{Including ChatGPT}  &     2 & 0 & 1 & 0 & 0 & 1 & 0 & 1 &  17 &  1 & 1 & 1 & 1 &  1  & 10 &6 & 20    \\
        &         &   & \multicolumn{3}{c}{Excluding ChatGPT}  &   2 & 0 & 1 & 0  & 0 & 5 & 0 & - &  - & 1 & 2 & 5 & 4  &  2  & 10 &7  &24    \\
    \bottomrule[1pt]
    \end{tabular}
     \label{tab:res}
\end{table*}

\subsection*{F. Examples of Query Response in Various LLMs}
\label{sec:appendixllm}

Table \ref{tab:llms} presents the examples of query and response among various LLMs on FinQA, FinTS, FinRE, FinED, FinER and FinNC task.

\begin{table*}[htb!]
    \centering
    \footnotesize
      \setlength\tabcolsep{1pt} 
       \renewcommand{\arraystretch}{1.5}
      \caption{Examples of query and response among various LLMs on some specific task of Question Answering(FinQA), Text Summarization(FinTS), Relationship Extraction(FinRE), Event Detection(FinED), Entity Recognition(FinER) and News Classification(FinNC).}
     \resizebox{\textwidth}{!}{%
    \begin{tabular}{c|m{8cm}|m{3cm}|m{4cm}|m{3cm}}
      \toprule[1pt]
        \multirow{2}{*}{\textbf{Specific Task}} & \multirow{2}{*}{\textbf{Query Input}} & \multicolumn{3}{c}{\textbf{Model Response}} \\ 
         \cline{3-5}
          & & \multicolumn{1}{c|}{\textbf{Baichuan-7B}} & \multicolumn{1}{c|}{\textbf{GPT-4}} & \multicolumn{1}{c}{\textbf{ICE-INTERN-full-7B}} \\ 
        \hline
        \multirow{2}{*}{FinQA} &浦东建设公告，公司子公司上海市浦东新区建设（集团）有限公司、上海浦兴路桥建设工程有限公司中标多项重大工程项目，中标金额总计为10.94亿元。\textcolor{blue}{请问中标事件对应的主体有哪些?}   &  浦东建设、浦兴路桥、浦东建设(集团) &  中标事件对应的主体有上海市浦东新区建设（集团）有限公司 & 上海市浦东新区建设（集团）有限公司,上海浦兴路桥建设工程有限公司   \\
        & Pudong Construction announced that the company's subsidiaries, Shanghai Pudong New Area Construction (Group) Company Limited and Shanghai Puxing Road and Bridge Construction Engineering Company Limited, have won a number of major projects, with a total winning amount of 1.094 billion yuan. \textcolor{blue}{What are the subjects corresponding to the winning bidding events?} & Pudong Construction, Puxing Road \& Bridge, Pudong Construction (Group) & The winning event corresponds to the subject of Shanghai Pudong New Area Construction (Group) Co. & Shanghai Pudong New Area Construction (Group) Company Limited,Shanghai Puxing Road \& Bridge Construction Engineering Co.    \\
        \hline
        \multirow{2}{*}{FinTS} &市场普遍预期日本央行将维持其超宽松货币政策不变，但知情人士表示，周二将公布的经济展望报告中，可能上调4月起财政年度的通胀预估。能源和原材料成本的上涨已将批发价格推高至40年来的最高水平，企业已开始将上涨的成本转嫁给消费者。近期通胀温和上升，主要是投入成本上升和日元疲软推动的，而不是强劲的国内需求。日本央行行长黑田东彦曾表示，物价上涨应该伴随着工资的强劲增长。\textcolor{blue}{请对上述金融报道进行总结，给出具有中心主旨的摘要，回答要求尽量简短。}  &日本央行行长黑田东彦曾表示，物价上涨应该伴随着工资的强劲增长。 &日本央行预计将上调4月起的通胀预估，尽管市场普遍预期货币政策将保持超宽松。成本上涨推高批发价格，企业已将涨成本转嫁给消费者，但通胀主要由投入成本上升和日元疲软推动，而非国内需求强劲。 &知情人士：日本央行可能上调4月起财政年度的通胀预估。\\
        & The Bank of Japan is widely expected to leave its ultra-loose monetary policy unchanged, but people familiar with the matter said it may raise its inflation forecast for the fiscal year starting in April in its economic outlook report to be released on Tuesday. Rising energy and raw material costs have pushed wholesale prices to the highest level in 40 years, and companies have begun to pass on rising costs to consumers. The recent moderate rise in inflation has been driven mainly by higher input costs and a weaker yen, rather than strong domestic demand. Bank of Japan Governor Haruhiko Kuroda has said price increases should be accompanied by strong wage growth. \textcolor{blue}{Please summarize the above financial reports, giving a summary with the central thrust, and keep your answers as brief as possible.} & Bank of Japan Governor Haruhiko Kuroda has said that price increases should be accompanied by strong wage growth.& The Bank of Japan is expected to raise its inflation forecast from April, although the market generally expects monetary policy to remain ultra-loose. Rising costs have pushed up wholesale prices and companies have passed on the cost increases to consumers, but inflation has been driven mainly by higher input costs and a weaker yen rather than strong domestic demand.&BOJ may raise inflation forecast for fiscal year from April: people familiar with the matter.    \\
        \hline
        \multirow{2}{*}{FinRE} &新希望集团销售额约为上市公司<N>倍有余,由于背靠农业帝国“希望系”, 资本市场一直预期集团能够向新希望股份注入农业资产, 如果集团资产整合启动, 新希望股份的股。\textcolor{blue}{请分析上述涉及实体对(新希望股份,新希望集团)的关系。}   &被收购  &被注资 &自己 \\
        & New Hope Group's sales are more than twice that of the listed company<N>, due to the backing of the agricultural empire "Hope", the capital market has been expecting the group to inject agricultural assets into New Hope shares, if the group's asset integration is launched, the shares of New Hope shares. \textcolor{blue}{Please analyze the relationship between the entities involved in the above (New Hope, New Hope Group).} & Acquired& Funded& Self    \\
        \hline
        \multirow{2}{*}{FinED} &大智慧(601519)信披违规遭调查“互联网+”炒作或终结金鹰等多家基金公司高管离职 基民担忧业绩波动。\textcolor{blue}{请分析上述金融报道中的事件类型及及其事件主体，并以‘事件类型，事件主体’的格式输出。}  &套现约1.5亿元&交易违规, 大智慧 &信批违规, 大智慧 \\
        & Great wisdom (601519) letter disclosure violations investigated "Internet +" speculation or the end of the Golden Eagle and many other fund company executives leave the funders worried about performance fluctuations. \textcolor{blue}{Please analyze the types of events and their event subjects in the above financial reports and output them in the format of 'event type, event subject'}. & Cashing in about \$150 million& Trading Violations, Great Intelligence &Letter Approval Violation, Great Intelligence    \\
        \hline
        \multirow{2}{*}{FinER} &本公司非执行董事孙月英女士不再担任中国远洋海运集团有限公司总会计师和中远财务有限责任公司董事长。\textcolor{blue}{请给出上述金融报道中存在的个人(’PER’)、组织(’ORG’)或地点(’LOC’)的特定命名实体，回答应遵循的格式’实体名称, 实体类型’。} & 中国远洋海运集团有限公司，ORG & 中远海运发展股份有限公司, ORG & 孙月英, PER; 中国远洋海运集团有限公司, ORG; 中远财务有限责任公司, ORG  \\
        & Ms. Sun Yueying, a non-executive director of the Company, ceased to be the chief accountant of China Ocean Shipping Group Company Limited and the chairman of COSCO Finance Company Limited. \textcolor{blue}{Please give the name of the person ('PER'), organization ('ORG') or location ('LOC') of the particular named entity that exists in the above financial report. ) of a specific named entity, the response should follow the format 'Entity Name, Entity Type'.} & China Ocean Shipping Group Company Limited, ORG& COSCO Shipping Development Company Limited, ORG& Sun Yueying, PER; China Ocean Shipping Group Company Limited, ORG; COSCO Finance LLC, ORG.    \\
        \hline
        \multirow{2}{*}{FinNC} &WTI原油涨幅回升至0.5\%，现报75.58美元/桶。\textcolor{blue}{请对这个金融新闻报道进行分类，具体属于['中国','外国','国际','公司','行业','大盘','经济','政策','政治','期货','债券','房地产','外汇','虚拟货币','新冠','能源']中的哪些类别？。} & \textcolor{red}{输出与内容无关} & 国际,能源 & 国际期货 \\
        & WTI crude oil has rallied to 0.5\% and is now at \$75.58 per barrel. \textcolor{blue}{Please categorize this financial news report, specifically which categories in ['China','Foreign','International','Companies','Industries','Broader Markets','Economy','Policies','Politics','Futures','Bonds','Real Estate','Forex','Virtual Currencies','New Crowns','Energy']}. & \textcolor{red}{output independent of content}& International, Energy&International futures    \\
        \bottomrule[1pt]
    \end{tabular}%
    }
    \label{tab:llms}
\end{table*}

\end{CJK}
\end{document}